\documentclass[12pt]{amsart}

\usepackage[dvips]{graphics}

\usepackage{epsfig}

\begin{document}

\title[]{\bf Conformal Anomalies for Interacting
             Scalar Fields on Curved Manifolds
             with Boundary}

\author[]{George Tsoupros \\
       {\em The School of Physics,}\\
       {\em Peking University,}\\
       {\em Beijing 100871,}\\
       {\em People's Republic of China}
}
\thanks{present e-mail address: gts@pku.edu.cn}

\begin{abstract}

The trace anomaly for a conformally invariant scalar field theory on a curved manifold of
positive constant curvature with boundary is considered. In the context of a perturbative
evaluation of the theory's effective action explicit calculations are given for those
contributions to the conformal anomaly which emerge as a result of free scalar propagation
as well as from scalar self-interactions up to second order in the scalar self-coupling.
The renormalisation-group behaviour of the theory is, subsequently, exploited in order to
advance the evaluation of the conformal anomaly to third order in the scalar self-coupling.
As a direct consequence the effective action is evaluated to the same order. In effect,
complete contributions to the theory's conformal anomaly and effective action are evaluated
up to fourth-loop order.


\end{abstract}

\maketitle


{\bf I. Introduction}\\

The fundamental physical significance of bounded manifolds has
been amply demonstrated in the framework of Euclidean Quantum
Gravity and, more recently, in the context of the holograpic
principle and the $ AdS/CFT$ correspondence. An issue of immediate
importance on such manifolds is the evaluation of the effective
action and, for that matter, of the conformal anomaly relevant to
the dynamical behaviour of quantised fields. This issue has
hitherto received attention, almost exclusively, at one-loop level
in the absence of matter-to-matter related interactions.

The present work extends the exploration of the dynamical
behaviour of scalar fields beyond one-loop order \cite{G},
\cite{T}, \cite{GT}, \cite{George} to the study of higher
loop-order contributions to the conformal anomaly for a
self-interacting massless scalar field conformally coupled to the
background geometry of the bounded manifold. Such an issue has
been studied on a general manifold without boundary
\cite{Collins}, \cite{Birrel}, \cite{OBS} as well as in the
specific case of de Sitter space \cite{DowkerCr},
\cite{DrummondShore}. In addition to the trace anomaly emerging at
one-loop order as a result of the background curvature and
presence of the boundary there is a further anomalous contribution
stemming from the self-interaction of the scalar field. Pivotal to
such an analysis is the perturbative evaluation of the
gravitational component of the renormalised effective action that
is, of the renormalised vacuum effective action. The stated
additional contributions which that component of the effective
action  perturbatively generates to the conformal anomaly are
closely associated with the renormalisation-group behaviour of the
theory. The structure of the renormalisation-group functions on a
general curved manifold with boundary has been discussed in
\cite{Od}, \cite{Odintsov}. The results hitherto obtained
\cite{G}, \cite{T}, \cite{GT}, \cite{George} allow for an explicit
calculation of the scalar-interaction-related contributions to the
anomalous trace up to second order in the scalar self-coupling. On
the basis of these results, the renormalisation group will,
subsequently, be exploited in order to attain an extension of such
an evaluation to order three in the scalar self-coupling.
Specifically, it will be shown through use of dimensional analysis
that to that order the anomalous contribution which emerges
exclusively as a result of scalar self-interactions is
proportional to the renormalisation-group function for the scalar
self-coupling in, both, the interior and boundary of the manifold.
In effect, the conformal anomaly will be successively evaluated to
first, second, third and fourth loop-order. Such an evaluation,
directly, reveals information about the renormalised vacuum
effective action at four-loop order. Its exact form at that order
can be elicited without use of a renormalisation procedure. As was
also the case with the renormalisation of the theory such an
evaluation will be simultaneously accomplished in the interior and
the boundary of the manifold order by order in perturbation.

For reasons of technical convenience the analysis will be
performed on $ C_4$, a segment of the Euclidean sphere bounded by
a hypersurface of positive extrinsic curvature, with homogeneous
Dirichlet-type boundary conditions. Such a choice allows for a
direct use of the results hitherto attained on such a manifold
\cite{G}, \cite{T}, \cite{GT}, \cite{George}. The results obtained
herein have a general significance for bounded manifolds of the
same topology both in terms of the general structure of the
effective action and in terms of the interaction between boundary
and surface terms. Notwithstanding that, such results as are
obtained herein on $ C_4$ deserve attention in their own merit due
to their additional significance for the Hartle-Hawking approach
to the quantisation of closed cosmological models.

As expected on the basis of the results already obtained, the
perturbative evaluation of the theory's conformal anomaly involves
substantially complicated expressions which render the associated
calculations arduous. However, as is also the case with the
theory's perturbative renormalisation, the underlying premise
which allows for the evaluation of the conformal anomaly is based
on the conceptually simple techniques which were initiated and
developed in \cite{G}, \cite{T}.

The contribution to the conformal anomaly which emerges from free
propagation on a curved manifold is the exclusive result of the
gravitational backreaction on the manifold's geometry and has a
distinct character from that which emerges from matter-to-matter
interactions. For that matter, the analysis in the context of free
scalar propagation in Sec.III is independent from the ensuing
analysis which pursues the scalar self-coupling-related
contributions to the conformal anomaly. A brief outline of the
techniques relevant to the evaluation of the
free-propagation-related component of the conformal anomaly on a
general bounded manifold is presented in Sec.III as the incipient
point of an analysis which advances from the general to the
concrete case of the bounded manifold of positive constant
curvature stated herein. Appendix II amounts, mostly, to a
citation of results directly relevant to the evaluation of the
trace anomaly. However, a study of the analysis relevant to a
certain mathematical aspect of the theory in Appendix I and in the
following segment of Appendix II is recommended prior to that of
sections IV and V.

{\bf II. Effective Action and Trace Anomaly on $ C_4$}\\

The scalar component of the bare action defining a theory for a
conformal, massless field $ \Phi$ specified on $ C_n$ - a manifold
of positive constant curvature embedded in a $ (n+1)$-dimensional
Euclidean space with embedding radius $ a$ and bounded by a $
(n-1)$-sphere of positive constant extrinsic curvature $ K$
(diverging normals) - at $ n=4$ is \cite{George}

$$
S[\Phi_0] =
\int_{C}d^4\eta\big{[}\frac{1}{2}\Phi_0(\frac{L^2-\frac{1}{2}n(n-2)}{2a^2})\Phi_0
- \frac{\lambda_0}{4!}\Phi_0^4 - \frac{1}{2}[\xi_0 R +
\kappa_0K^2]\Phi_0^2 \big{]}
$$

\begin{equation}
+ \oint_{\partial C}d^3\eta(\epsilon_0K\Phi_0^2)
\end{equation}
with the subscript $ C$ signifying integration in the interior of
$ C_4$ and with the subscript $ \partial C$ signifying integration
exclusively on its boundary. In either case the subscript $ 4$ has
been omitted as the integration itself renders the associated
dimensionality manifest. In (1) $ \eta$ is the position vector in
the embedding $ (n + 1)$-dimensional Euclidean space signifying
the coordinates $ \eta_{\mu}$ and

$$
L_{\mu \nu} = \eta_{\mu}\frac{\partial}{\partial \eta_{\nu}} -
\eta_{\nu}\frac{\partial}{\partial \eta_{\mu}}
$$
is the generator of rotations. In addition, the Ricci scalar $ R$ relates to the constant embedding radius $ a$ through

\begin{equation}
R = \frac{n(n-1)}{a^2}
\end{equation}
As stated, the bare action (1) is associated by choice with the homogeneous Dirichlet condition $ \Phi_{|\partial C_4}=0$ for the scalar field.

The gravitational component of the bare action on $ C_n$ at $n \rightarrow 4$ is

$$
S_g = \int_Cd^4\eta\big{[}\frac{1}{8\pi G_0}\Lambda_0 - \frac{1}{16\pi G_0}(R + K^2) + \alpha_0R^2 + \theta_0RK^2\big{]}$$

\begin{equation}
+ \oint_{\partial C}d^3\eta\big{[}\frac{(-1)}{8\pi G_0}
\sigma_0K + \zeta_0RK + \delta_0K^3\big{]}
\end{equation}
where with the exception of $ G_0$ and $ \Lambda_0$ all other bare parameters have no mass dimensions.

All correlation functions in the context of a quantum field theory emerge through functional differentiation of the vacuum-to-vacuum amplitude with respect to a classical source $ J$. If all fields are quantised on a fixed background manifold the Einstein field equations necessitate the mean value of the stress energy-momentum tensor $ T_{\mu\nu}$ in any quantum state to be independent from that source. For that matter, the mean value for the stress energy-momentum tensor of a quantum field emerges, itself, as a result of a functional differentiation of the vacuum-to-vacuum amplitude with respect to the background metric, at $ J=0$. Specifically, with $ W[g_{\mu\nu}; J]$ being the generating functional for connected Green functions, it is

$$
<0,out|T_{\mu\nu}|0,in> = \frac{2}{\sqrt{-g}}\frac{\delta}{\delta g^{\mu\nu}}W[g_{\mu\nu}; J=0]
$$
As $ W[g_{\mu\nu}; J=0]$ can be seen, through a functional Legendre transformation, to be in  formal coincidence with the gravitational component $ W$ of the theory's effective action on a curved manifold it follows that the vacuum expectation value of the stress tensor amounts perturbatively to the sum-total of diagramatic vacuum contributions at each order. The diagrams relevant to such contributions are, necessarily, characterised by the absence of external wavefunctions. It is worth mentioning that a significant advantage of this approach to the renormalisation of the stress tensor is that the vacuum states need not be specified \cite{Birrel}. Fig.1 lists all diagrams relevant to the perturbative evaluation of the gravitational effective action $ W$ up to third loop-order. It will be shown, in what follows, that up to that order these diagrams are associated with the perturbative evaluation of the zero-point function $ \Gamma^{(0)}$ and, by extension, with the vacuum effects responsible for the back-reaction of the scalar field on the background geometry of $ C_4$. Perturbatively, they are also, for that matter, the exclusive source of the theory's trace anomaly to which they signify contributions independent of the state of the quantum scalar field.

\begin{figure}[ht]
\centering\epsfig{figure=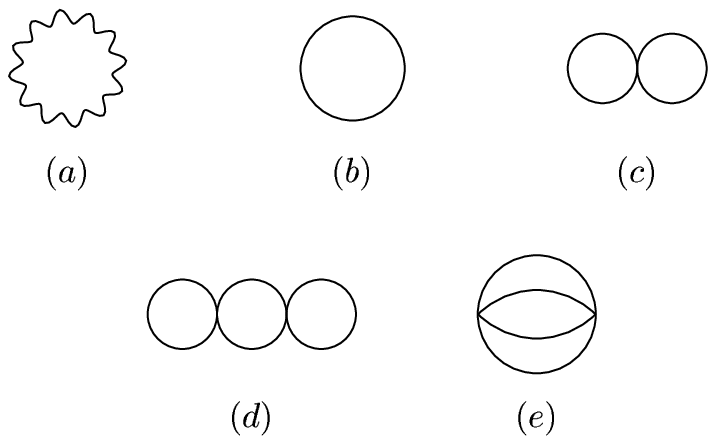, height = 60mm,width=100mm}
\caption{Diagrams contributing to $ O(\lambda^2)$-related vacuum effective action.}
\end{figure}
The boundary term in (3) whose dependence on geometry is featured exclusively on $ K$
is essentially the Gibbons-Hawking term in the gravitational action functional and is irrelevant to the renormalisation of the theory beyond one-loop order \cite{George}. At one-loop, however, it signifies a quantum correction to the Einstein-Hilbert action which deserves attention in its own merit since it emerges as the result of the influence which the background curvature and boundary have on free propagation. The Gibbons-Hawking term in (3) is the boundary component of the sum-total of contributions represented by fig.1(a) and fig.1(b). These diagrams are distinct from all other as in the loop expansion they correspond to contributions independent of the embedding radius $ a$. At one-loop order the graviton contribution in fig.1(a) to the bare gravitational action is of equal importance as that effected by vacuum scalar effects in fig.1(b) \cite{Birrel}. The sum-total of their contributions results, through renormalisation, in the one-loop effective action associated with free propagation. Such contributions to the bare gravitational action on a general manifold with boundary as are represented here by the diagrams in fig.1(a) and fig.1(b) have been the object of extensive study in Euclidean Quantum Gravity through use of zeta-function techniques and heat kernel asymptotic expansions \cite{EKP}. In the context of the theory pursued herein  the one-loop effective action and associated trace anomaly can be obtained from these general results on a bounded manifold $ M$ by specifying the geometry to be that of $ C_4$ with homogeneous Dirichlet conditions on $ \partial C_4$ and the coupling between matter and gravity to be a conformal coupling between a scalar field and the stated geometry. The outline of the associated calculation leading to the trace anomaly follows.

{\bf III. Trace Anomaly and Free Scalar Propagation on $ C_4$}

The one-loop effective action $ W_0$ associated with the free scalar propagation in fig.1(b) on a general manifold $ M$ is, generally, given by the expression

\begin{equation}
W_0 = \frac{1}{2}TrlnD
\end{equation}
where $ D$ is the operator associated with the scalar propagator on $ M$, acting on an abstract Hilbert space of states $ |n>$ subject to orthonormality conditions with eigenvalues $ \lambda_n$ \cite{Birrel}. Introducing the generalised $ \zeta$-function as \cite{EKP}

\begin{equation}
\zeta(s) \equiv Tr[D^{-s}] = \sum_{n}\lambda_n^{-s}
\end{equation}
it follows that in the case of the scalar field it is

\begin{equation}
W_0 = -\frac{1}{2}\frac{\partial}{\partial s}\zeta(s)_{|s=0} - \frac{1}{2}\zeta(0)ln(\mu^2)
\end{equation}
On the grounds of general theoretical considerations outlined in the previous paragraph the mean value of the stress energy-momentum tensor in some vacuum state is

\begin{equation}
<T_{\mu\nu}> = \frac{2}{\sqrt{-g}}\frac{\delta W}{\delta g^{\mu\nu}}
\end{equation}
As the gravitational effective action $ W$ on $ M$ amounts perturbatively to a sum over contributions expressed by vacuum diagrams such as those in fig.1 it admits the form

\begin{equation}
W = W_0 + W_I
\end{equation}
with $ W_I$ being the component of $ W$ perturbatively generated by vacuum diagrams with interaction vertices such as the last three diagrams in fig.1.

Such counterterms contained perturbatively in the bare gravitational action as are necessary to cancel the divergences which appear in the $ W_0$ component of $ W$ on a general manifold $ M$ are local in the metric field and conformally invariant in four dimensions. In effect, the trace of the renormalised stress tensor in (7) receives a contribution from $ W_0$ which can be seen from (6) to relate to $ \zeta(s)$ through \cite{EKP}

\begin{equation}
\int_Md^4x\sqrt{-g}<T^c_c>_r = \zeta(0)
\end{equation}
On a general manifold $ M$, for that matter, the conformal anomaly emerging from free propagation at one-loop level for conformally invariant theories is specified by $ \zeta(0)$. This result remains valid in the presence of a non-trivial boundary on the understanding that integration over $ M$ also includes $ \partial M$.

In order to evaluate the trace anomaly associated with free propagation, such as that inherent in fig.1b, on $ M$ it is necessary to consider the asymptotic expansion \cite{EKP}

\begin{equation}
G(t) \sim \sum_{k=0}^{\infty}A_{k}t^{\frac{k - 4}{2}}; ~~~t \rightarrow 0^{+}
\end{equation}
of the supertrace

\begin{equation}
G(t) = \int_CtrK_D(x, x,t)d^4x = \sum_ne^{-\lambda_nt} = Tre^{-tD}
\end{equation}
of the heat kernel

$$ K_D(x, x',t) = \sum_n<n|x'><x|n>e^{-\lambda_nt} $$
associated with the bounded elliptic operator $ D$ in (4) through the heat equation

\begin{equation}
(\frac{\partial}{\partial t} + D)K_D(x, x',t) = 0
\end{equation}
The supertrace relates to $ \zeta(s)$ through an inverse Mellin transform as

\begin{equation}
\zeta(s) = \frac{1}{\Gamma(s)}\int_0^{\infty}t^{s-1}G(t)dt
\end{equation}
The asymptotic expansion in (10) yields, in the context of (13), the result

\begin{equation}
\zeta(0) = A_4
\end{equation}
which, as (9) reveals, reduces the issue of the conformal anomaly due to free propagation of matter on $ M$ to the issue of evaluating the constant coefficient $ A_4$ in (10).

The general asymptotic expansion in (10) is characterised by the exclusive presence of even-order coefficients $ A_{2k}$ on any manifold $ M$ for which $ \partial M=0$. The presence of a non-trivial $ \partial M$ has the effect of generating an additional boundary-related component for each even-order coefficient as well as non-vanishing boundary-related values for all $ A_{2k+1}$ in (10). In general, the coefficients for the supertrace of the heat kernel associated with the relevant elliptic operator on a bounded four-dimensional manifold $ M$ admit, in the context of (10), the form

\begin{equation}
A_{2k} = \int_Ma^{(0)}_{2k}\sqrt{g}d^4x + \int_{\partial M}a^{(1)}_{2k}\sqrt{h}d^3x
\end{equation}

\begin{equation}
A_{2k+1} = \int_{\partial M}a^{(1)}_{2k+1}\sqrt{h}d^3x
\end{equation}
with $ h$ being the induced metric on the boundary.

The local interior coefficients $ a^{(0)}_{2k}$ are specified by the same local invariants as in the unbounded manifold of the same local geometry and do not, for that matter, depend on the boundary conditions. The far more complicated boundary coefficients $ a^{(1)}_k$ necessitate knowledge of the geometry of $ \partial M$ and of the boundary conditions on it in addition to knowledge of the geometry of $ M$.

If $ M$ is specified to be a Riemannian manifold of positive constant curvature then $ M$ reduces to $ S_4$ in the absence of a boundary and to $ C_4$ if $ \partial C_4$ is, itself, specified to be a Euclidean three-sphere of constant extrinsic cuvature. In either case, the elliptic operator $ D$ in (4) associated with free propagation of a massless scalar field conformally coupled to the background metric is the operator which appears as the d'Alembertian in (1)

\begin{equation}
D = \frac{L^2-\frac{1}{2}n(n-2)}{2a^2}
\end{equation}
This elliptic operator is unbounded on the Euclidean de Sitter space $ S_4$ and
its zeta-function evaluation results in the one-loop effective action \cite{Collins},
\cite{DrummondShore}

\begin{equation}
W_0 = \frac{1}{90}\frac{1}{\epsilon} + O(\epsilon^0);~~~ \epsilon = 4 - n
\end{equation}
for a conformal scalar field with an associated anomalous trace contribution

\begin{equation}
<T^c_c>_r = -\frac{1}{90}\frac{1}{a^4\Omega_5}
\end{equation}

The elliptic operator in (17) is bounded on $ C_4$. The evaluation of the trace anomaly due to free scalar propagation on that manifold is associated with fig.1b and necessitates the asymptotic expansion (10) of the supertrace

\begin{equation}
G(t) = \int_CtrK_D(\eta,\eta,t)d^4\eta
\end{equation}
of the heat kernel $ K_D(\eta,\eta',t)$ associated with the bounded elliptic operator $ D$ in (17).

It is worth emphasising that, despite appearances stemming from the homogeneous Dirichlet condition $ \Phi_{|\partial C_4}=0$, the boundary term

$$
\oint_{\partial C}d^3\eta(\epsilon_0K\Phi_0^2)$$
in (1) does not vanish. Such a non-vanishing effect arises as a result of the boundary condition

\begin{equation}
K_D(\eta,\eta',t=0) = \delta^{(4)}(\eta-\eta')
\end{equation}
- imposed on the solution to the heat equation on $ C_4$

\begin{equation}
(\frac{\partial}{\partial t} + D)K_D(\eta,\eta',t) = 0
\end{equation}
- which offsets the effect of the homogeneous Dirichlet condition
on $ \partial C_4$ \cite{BarvSol}.

The expressions (15) and (16) for the expansion coefficients $ A_k$ in (10) reduce, respectively, to

\begin{equation}
A_{2k}(D,C_4) = \int_Cc^{(0)}_{2k}d^4\eta + \int_{\partial C}c^{(1)}_{2k}d^3\eta_{B}
\end{equation}
and

\begin{equation}
A_{2k+1}(D,C_4) = \int_{\partial C}c^{(1)}_{2k+1}d^3\eta_B
\end{equation}
on $C_4$, with $ D$ specified in (17) and with the embedding coordinate vector $ \eta_{B}$ specifying the spherical boundary hypersurface of maximum colatitude $ \theta_0$. As stated in the context of (9) and (14), the trace anomaly on $ C_4$ is associated with the $ A_4(D,C_4)$ coefficient in (23).

If the curvature of a Riemannian manifold $ M$ satisfies the vacuum Einstein equations with a cosmological constant

\begin{equation}
R_{\mu\nu} = \Lambda g_{\mu\nu}
\end{equation}
then the coefficients $ a^{(0)}_{4}$ and $ a^{(1)}_{4}$ specifying $ A_4$ in (15) will admit the expressions \cite{MossPol}

\begin{equation}
a^{(0)}_{4} = \alpha_0\Lambda^2 + \alpha_2R_{abcd}R^{abcd}
\end{equation}
and

\begin{equation}
a^{(1)}_{4} = \beta_1\Lambda k + \beta_2 k^3 + \beta_3 kk_{ab}k^{ab} + \beta_4k_a^bk_b^ck_c^a + \beta_5C_{abcd}k^{ac}n^bn^d
\end{equation}
respectively, where in (27) $ k_{ab}$ is the extrinsic curvature of $ \partial M$ and $ n$
is the vector normal to $ \partial M$. The expressions in (26) and (27) essentially disentangle the geometry-related contributions to the $ A_4$ coefficient from those contributions which depend on the elliptic operator and boundary conditions. Specifically, the coefficients $ \alpha_0$ and $ \alpha_2$ multiplying respectively the geometry-dependent expressions in (26) depend exclusively on the operator in whose heat-kernel asymptotic expansion $ A_4$ is the constant coefficient. Likewise, the five coefficients $ \beta_i$ which respectively multiply the geometry-related expressions in (27) depend only on the same operator and the conditions specified on the boundary.

If, in the case of $ \Lambda > 0$, a boundary condition imposed on (25) is that of a compact four-geometry then the solution to (25) can be either the spherical cap $ C_4$ or the Euclidean four-sphere $ S_4$. The former case emerges if the remaining boundary condition corresponds to the specification of the induced three-geometry as a three-sphere. The latter case emerges if the remaining boundary condition corresponds to the absence of a boundary. In addition, the former case reduces to a disk $ \mathcal{D}$ at the limit of boundary three-spheres small enough to allow for their embedding in flat Euclidean four-space. These three solutions are aspects of the Hartle-Hawking no-boundary proposal for the quantisation of closed universes \cite{HarHawk}. For the stated boundary conditions these solutions to (25) also coincide with the corresponding solutions to the Euclidean Einstein field equations in the presence of a massless scalar field conformally coupled to gravity on the additional Dirichlet condition of a constant field on $ \partial C_4$, in the present case of $ C_4$ as well as in that of $ \mathcal{D}$. Such a coincidence is a consequence of a vanishing stress tensor for the conformal scalar field \cite{HarHawk}.

In effect, the constant coefficient $ A_4(D,C_4)$ in the heat kernel asymptotic expansion for a conformal scalar field on $ C_4$, the corresponding $ A_4(D,S)$ on $ S_4$, as well as the corresponding coefficient $ A_4(D,\mathcal{D})$ on $ \mathcal{D}$ are expected to be inherently related. With $ \theta_0$ being the maximum colatitude on $ C_4$, which for that matter specifies $ \partial C_4$, the stated relation is \cite{MossPol}

\begin{equation}
A_4(D,C_4) = A_4(D,S)(\frac{1}{2} - \frac{3}{4}cos\theta_0 + \frac{1}{4}cos^3\theta_0) + A_4(D,\mathcal{D})cos^3\theta_0 + \ \frac{9}{8}\beta_1cos\theta_0sin^2\theta_0
\end{equation}
where, in conformity with (9), (14) and (19), it is

\begin{equation}
A_4(D,S) = -\frac{1}{90}
\end{equation}
Moreover, the corresponding value in the case of $ \Phi_{|\partial\mathcal{D}}=0$ is \cite{EKP}

\begin{equation}
A_4(D,\mathcal{D}) = - \frac{1}{180}
\end{equation}
and the value of the coefficient $ \beta_1$ for the present case of $ C_4$ with $ \Phi_{|\partial C_4}=0$ is \cite{MossPol}

\begin{equation}
\beta_1 = \frac{29}{135}
\end{equation}
In effect, equation (28) yields $ A_4(D,C_4)$ for the free scalar propagation in fig.1b. At one-loop fig.1a also yields a contribution to the gravitational effective action which, as stated, is as important as the contribution which corresponds to fig.1b. The associated $ A_4$ coefficient for absolute boundary conditions can be evaluated through the, corresponding to (28), relation which includes the additional ghost contributions \cite{MossPol}. As the objective herein is the conformal anomaly generated by vacuum scalar effects such a contribution will be examined no further.

In the context of (9) and (14) the result which (28)-(31) signify relates to the conformal anomaly through

\begin{equation}
\int_C d^4\eta <T^c_c>_r^{(C)} + \int_{\partial C} d^3\eta <T^c_c>_r^{(\partial C)} = A_4(D, C_4)
\end{equation}

In order to arrive at a local expression for the trace anomaly on
$ C_4$ use will be made of the stated fact that on any bounded
manifold the local interior coefficients $ a_{2k}^{(0)}$,
associated through (15) with the asymptotic expansion of the
supertrace of the heat kernel in (10), are specified by the same
local invariants as in the unbounded manifold of the same local
geometry. This, in turn, reveals in the context of (10) that the
local asymptotic expansion of the heat kernel associated with the
operator $ D$ in (17) exclusively in the interior of $ C_4$ is in
coincidence with the local asymptotic expansion of the heat kernel
for the same operator on $ S_4$ so that (23) yields

\begin{equation}
c_{2k}^{(0)} = s_{2k}^{(0)}
\end{equation}
with $ s_{2k}^{(0)}$ being the local coefficients $ a_{2k}^{(0)}$ if $ M$ in (15) is specified as $ S_4$. Setting $ k=2$ and integrating in the interior of $ C_4$ yields

\begin{equation}
\int_Cd^4\eta c_{4}^{(0)} = \int_Cd^4\eta s_{4}^{(0)}
\end{equation}
and, through (32) and (23)

\begin{equation}
<T_c^c>_r^{(C)}  = s_{4}^{(0)}
\end{equation}
In the context of (15), however, (19) amounts to

\begin{equation}
\int_Sd^4\eta <T_c^c>_r  = -\frac{1}{90} = \int_Sd^4\eta s_{4}^{(0)}
\end{equation}
so that in view of the constancy of $ s_{2k}^{(0)}$ on $ S_4$ it is

\begin{equation}
s_{4}^{(0)} = -\frac{1}{90}\frac{1}{a^4\Omega_5}
\end{equation}
Substituting this result in (35) yields

\begin{equation}
<T_c^c>_r^{(C)}  = -\frac{1}{90}\frac{1}{a^4\Omega_5}
\end{equation}
This is the desired local expression for the trace anomaly in the interior of $ C_4$. As expected, it coincides with the corresponding expression in (19) for the trace anomaly
on $ S_4$.

Finally, substituting (38) in (32) yields

\begin{equation}
\int_{\partial C} d^3\eta <T^c_c>_r^{(\partial C)} = A_4(D, C_4) +
\frac{1}{90}(-\frac{1}{3}sin^2\theta_4^0cos\theta_4^0 -
\frac{2}{3}cos\theta_4^0 + \frac{2}{3})
\end{equation}
with \cite{G}

\begin{equation}
\int_Cd^4\eta = a^42\pi^2(-\frac{1}{3}sin^2\theta_4^0cos\theta_4^0 -
\frac{2}{3}cos\theta_4^0 + \frac{2}{3})
\end{equation}
Again, the boundary-related contribution $ <T^c_c>_r^{(\partial C)}$ is constant on $ \partial C_4 \equiv S_3$. Consequently,

\begin{equation}
<T_c^c>_r^{(\partial C)} = \frac{1}{a^34\pi}A_4(D, C_4) +
\frac{1}{90}\frac{1}{a^34\pi}(-\frac{1}{3}sin^2\theta_4^0cos\theta_4^0 -
\frac{2}{3}cos\theta_4^0 + \frac{2}{3})
\end{equation}
This is the desired local expression for the trace anomaly on $ \partial C_4$. As expected, it is contingent upon the specified homogeneous Dirichlet condition through (28)-(31).

{\bf IV. Trace Anomaly and Interacting Scalar Fields on $ C_4$}

The contribution to the trace anomaly stemming at order one in the scalar self-coupling is associated with the diagram of fig.1(c) and is, obviously, finite since at that order the latter signifies a finite contribution to the effective action \cite{George}. It will be shown to be vanishing in the context of the ensuing calculation. The evaluation of higher loop-order contributions to the trace anomaly on $ C_4$ pursued herein commences at $ O(\lambda^2)$ and will be the result of explicit diagramatic calculations. All vacuum scalar contributions to the gravitational effective action up to second order in the scalar self-coupling are represented diagramatically by the diagrams in fig.1(b), fig.1(c), fig. 1(d) and fig.1(e) respectively. Of them, the last three represent contributions which involve the scalar self-interactions. They amount to contributions of up to third loop-order. The divergences which such diagrams entail are cancelled by counterterms generated by the renormalisation process up to the stated $ O(\lambda^2)$. As was shown in \cite{George} the diagramatic structure in fig.1(c) entails no primitive divergences and the counterterm associated with that diagram's overlapping divergence arises from the replacement of the bare coupling constant $ \lambda_0$ by its expansion

\begin{equation}
\lambda_0 = \mu^{\epsilon}\big{[}\lambda + \sum_{\nu=1}^{\infty}\frac{a_{\nu}(\lambda)}{\epsilon^{\nu}}\big{]} =
\mu^{\epsilon}\big{[}\lambda + \sum_{k=1}^{\infty}\sum_{i=k}^{\infty}\frac{a_{ki}\lambda^i}{\epsilon^{k}}\big{]}
\end{equation}
in powers of $ \lambda$. In fact, counterterms stemming from such a replacement at the dimensional limit $ \epsilon \equiv (4-n) \rightarrow 0$ are expected to any order above one-loop level in the renormalisation process.

To $ O(\lambda^2)$ the contribution of fig.1(c), fig.1(d) and fig.1(e) to the gravitational component of the effective action amounts to \cite{George}

$$
\lambda^2
\frac{1}{\pi^{4}}\frac{1}{3}\frac{1}{2^{12}}\big{[}\sum_{N=0}^{\infty}\big{(}
\frac{1}{\pi^{2}}\frac{1}{3^3}\frac{1}{2^7}\frac{(N+2)(N+3)}{(N+1)(N+4)}
\big{(}C_{N+1}^{\frac{3}{2}}(cos{\theta_4^0})\big{)}^2 + $$

$$
\sum_{N'=0}^{N'_0}\frac{1}{[{N'}^2-N^2+3(N'-N)]^2}[\Gamma(\frac{1}{N'_0}) +
\frac{1}{3}\frac{1}{2}\frac{\Gamma(1+\frac{1}{N'_0})\Gamma(N'+1+\frac{1}{N'_0})}{\Gamma(N'+3+\frac{1}{N'_0})}(N+1)(N+2)]\times $$

$$
\big{[}\frac{3}{2^3}\frac{1}{\pi^2}
F^2\big{]} -
\big{[}\frac{3}{2^{9}}\frac{(V_c + 1)^2}{\pi^5} + \frac{\pi^3}{2^{5}}(V_c + 1)\big{]}
\big{[}\sum_{N'=0}^{N'_0}(2N' + 3)\big{]}^2
\sum_{N=1}^{\infty}\frac{[C_N^{\frac{3}{2}}(cos{\theta_4^0})]^2}{N^2(N+3)^2}\big{]}\big{)}\big{]}RK^2\frac{1}{\epsilon}\int_Cd^4\eta +
$$

$$
\lambda^2\frac{1}{\pi^{6}}\frac{1}{3}\frac{1}{2^{17}}\sum_{N=0}^{\infty}\sum_{N'=0}^{N'_0}
\frac{1}{[{N'}^2-N^2+3(N'-N)]^2} \times $$

$$
\left[ \Gamma(\frac{1}{N'_0}) +
\frac{1}{3}\frac{1}{2}\frac{\Gamma(1+\frac{1}{N'_0})\Gamma(N'+1+\frac{1}{N'_0})}{\Gamma(N'+3+\frac{1}{N'_0})}(N+1)(N+2) \right](BH)^2
\frac{R^2}{\epsilon}\int_Cd^4{\eta} +  $$

$$
\lambda^2\frac{1}{\pi^{6}}\frac{1}{3}\frac{1}{2^{12}}\sum_{N=0}^{\infty}\sum_{N'=0}^{N'_0}
\frac{1}{[{N'}^2-N^2+3(N'-N)]^2} \times $$

$$
\left[ \Gamma(\frac{1}{N'_0}) +
\frac{1}{3}\frac{1}{2}\frac{\Gamma(1+\frac{1}{N'_0})\Gamma(N'+1+\frac{1}{N'_0})}{\Gamma(N'+3+\frac{1}{N'_0})}(N+1)(N+2) \right]\times $$

\begin{equation}
\hspace{1.5in}
(FBH)\big{(}sin(\theta_4^0)\big{)}^{-3}
\frac{RK}{\epsilon}\oint_{\partial C}d^3{\eta} + F. T.
\hspace{2.0in}
\end{equation}
where the finite terms remain to be assessed. In this expression $
N$ is the quantum number associated with the angular momentum
flowing through the singular part of the propagator and $ N'$ is
the corresponding quantum number associated with the boundary part
of the propagator. The exact nature of the transform-space cut-off
$ N'_0$ has been cited and analysed in \cite{G} and summarised in
Appendix I. The n-dependent quantities $ F, B$ and $ H$ are
complicated functions of Gegenbauer polynomials. They have been
derived in \cite{T}, \cite{GT} and reproduced in \cite{George}.
For the sake of completeness they are given below at $ n
\rightarrow 4$.

$$
F(4) =
\sum_{m_1=0}^{N'cos{\theta_4^0}}(sin{\theta_4^0})^{2m_1}C_{N-m_1}^{m_1+\frac{3}{2}}(cos{\theta_4^0})
C_{N'-m_1}^{m_1+\frac{3}{2}}(cos{\theta_4^0})m_1
\hspace{1.0in}
$$
with $ \theta_n^0$ being the angle specifying $ \partial C_n$ in the $ (n+1)$-dimensional embedding Euclidean space and with

$$
m_1^0 = Ncos{\theta_n^0}~~;~~l_1^0 = N'cos{\theta_n^0}
\hspace{2.0in}
$$
being the degrees of spherical harmonics defined on $ \partial C_n$.

$$
{B(4)} = 2\big{[}m_1(sin{\theta_4^0})^{m_{1}+l_{1}+2}(cos{\theta_4^0})
C_{N-m_{1}}^{m_1+\frac{3}{2}}(cos{\theta_4^0})~~ - $$

$$
\hspace{1.0in}
(sin{\theta_4^0})^{m_{1}+l_{1}+4}(2m_1+3)C_{N-m_{1}-1}^{m_1+\frac{5}{2}}(cos{\theta_4^0})\big{]}
C_{N'-l_{1}}^{l_1+\frac{3}{2}}(cos{\theta_4^0})
\hspace{3in}
$$

$$
H(4) = \prod_{k=1}^{2}\int_0^{\pi}C_{m_k-m_{k+1}}^{m_{k+1}+\frac{3}{2}-\frac{1}{2}k}(cos{\theta_{4-k}})
C_{l_k-l_{k+1}}^{l_{k+1}+\frac{3}{2}-\frac{1}{2}k}(cos{\theta_{4-k}})[sin{\theta_{4-k}}]^{m_{k+1}+l_{k+1}+1}
d{\theta_{4-k}}\times $$

$$
\hspace{2.0in}
\int_0^{2\pi}\frac{e^{i(\pm m_{3}\mp l_{3})\theta_1}}{sin{\theta_1}}d\theta_1
\hspace{2.5in}
$$

As the effective action amounts perturbatively to a summation over diagramatic contributions (43) yields the entire contribution to the, as of yet unrenormalised, gravitational effective action $ W$ to $ O(\lambda^2)$. In terms of the loop expansion (43) is, in the context of (8), the complete three-loop $ \big{(}O(\hbar^3)\big{)}$ contribution to $ W_I$.

The general relation between the gravitational effective action $ W$ and the conformal anomaly in the presence of matter-to-matter couplings has been outlined in the second section of this work. More precisely, it can be elicited from the generating functional $ W[g_{\mu\nu}; J]$ for connected Green functions

\begin{equation}
e^{W[g_{\mu\nu}; J]} = \int\mathcal{D}\Phi e^{S[\Phi]}
\end{equation}
The formal relation between $ W[g_{\mu\nu}; J]$ and the effective action through a functional Legendre transformation reveals that on a curved manifold the component of the effective action responsible for the generation of all back-reaction effects of the matter field on the metric is the vacuum effective action $ W[g_{\mu\nu};J=0] \equiv W$. The gravitational component $ W$ of the effective action determines perturbatively, for that matter, all contributions to, both, the vacuum expectation value of the stress tensor and the trace anomaly stemming from matter-to-matter interactions and Yukawa couplings in addition to that stemming from free propagation \cite{Birrel}, \cite{OBS}. Consequently, being perturbatively generated by diagramatic contributions which are characterised by the absence of external wavefunctions, $ W_I$ is the exclusive source of the additional component of the trace anomaly at loop-orders above one.

In the context of the present theory $ W[g_{\mu\nu}; J]$ is generated through a functional integration over the exponentiated action functional $ S[\Phi]$ formally associated with (1). If the functional derivative of $ W[g_{\mu\nu};J=0]$ is specified at that metric which corresponds to a manifold of positive constant curvature characterised by embedding radius $ a$ then the general relation (7) results in \cite{DrummondShore}

\begin{equation}
\int_Cd^4\eta <T^c_c> = a \frac{\partial W}{\partial a}
\end{equation}
which, by virtue of (8), reveals the relation between the component $ W_{I_r}$ of the renormalised gravitational effective action generated by vacuum diagrams with vertices and the higher loop-order contributions to the trace of the renormalised stress tensor $ <T^c_c>_{I_r}$ to be

\begin{equation}
\int_Cd^4\eta <T^c_c>_{I_r} = a \frac{\partial W_{I_r}}{\partial a}
\end{equation}
This expression relates $ <T^c_c>_{I_r}$ and $ W_{I_r}$ through the background geometry and can be seen to have both formal and perturbative significance.

The result of the perturbative evaluation of $ W$, expressed to $ O(\hbar^3)$ as $ W_I^{(3)}$ in (43), contains three divergences at the dimensional limit $ \epsilon \rightarrow 0$. They are all of order $ \lambda^2$. As shown in \cite{George} the $ RK^2$-related divergence cancels against the corresponding countertem contained in the bare gravitational coupling $ \theta_0$ in (3) to generate the $ \theta_{12}\lambda^2$ term in the latter's expansion in terms of $ \lambda$. Likewise, the $ R^2$-related divergence defines the corresponding $ O(\lambda^2)$ term in the relevant expansion for $ \alpha_0$ and the surface, $ RK$-related divergence, defines the corresponding $ O(\lambda^2)$ term in the relevant expansion for $ \zeta_0$. As $ W_{I_r}^{(3)}$ amounts to the finite terms in (43) it is imperative that the latter be assessed.

The assessment of the stated finite parts involves, in fact, the entire renormalisation program up to $ O(\lambda^2)$. As the evaluation of the relevant counterterms  necessitates only the divergent parts of the associated diagrams in the context of the minimal subtraction and mass-independent renormalisation the finite parts of those diagrams were not cited in \cite{George}. The assessment of those parts simply necessitates the expansion of all $ \epsilon$-dependent quantities in the diagramatic contributions about $ \epsilon=0$. In \cite{George} such an expansion was performed only in the relevant $ \Gamma$-functions for it is only such expansions which provide the divergent parts as poles in $ \epsilon$. The additional contributions to the stated finite parts will emerge from expansions of the form

\begin{equation}
(\mu a)^{\epsilon} = 1 + \epsilon ln(\mu a) + O(\epsilon^2)
\end{equation}
at the $ \epsilon \rightarrow 0$ limit in all radius-dependent terms. It is obvious that the pole structure of the theory remains intact in the context of such expansions on account of the presence of the $ \epsilon^0$-term. It should, in addition, be recalled that the renormalisation scheme of minimal subtraction, invoked herein, is characterised by
the perturbative absence of finite parts in all counterterms.

The results relevant to the finite terms of the vacuum effective action in (43) have been derived in Appendix II and can be placed in the context of the ensuing analysis simply by inspection. In addition, the mathematically important commutativity between the finite summations in these results and integration over $ C_4$ is the subject of Appendix I, a study of which is recommended. In effect, the entire three-loop contribution to the scalar-interaction-related sector of the renormalised gravitational effective action $ W_{I_r}^{(3)}$, which is also the entire contribution to $ O(\lambda^2)$, amounts to

\begin{equation}
W_{I_r}^{(3)} = (1c) + (1d) + (1e)
\end{equation}

In order to arrive at a local expression for the trace anomaly in both the interior and boundary of $ C_4$ the nature of $ W_{I_r}^{(3)}$ as the sum-total of contributions from both the interior and boundary, explicitly featured in (48), will be exploited in order to recast (46) in the form

\begin{equation}
\int_Cd^4\eta <T_c^c>_{I_r}^{(C)} + \oint_{\partial C}d^3\eta <T_c^c>_{I_r}^{(\partial C)} =
a\frac{\partial}{\partial a} \big{[}\int_Cd^4\eta L_C^{(3)}(\eta) + \oint_{\partial C}d^3\eta L_{\partial C}^{(3)}(\eta)\big{]}
\end{equation}
with $ L_C^{(3)}(\eta)$ and $ L_{\partial C}^{(3)}(\eta)$ being the interaction-related effective Lagrangian evaluated to $ O(\hbar^3)$ in the interior and boundary of $ C_4$ respectively. As partial differentiation with respect to the embedding radius $ a$, explicitly featured only in the logarithms of the Lagrangian, commutes with volume integration this yields

\begin{equation}
<T_c^c>_{I_r}^{(C)} =
a\frac{\partial}{\partial a}L_C^{(3)}
\end{equation}
and

\begin{equation}
<T_c^c>_{I_r}^{(\partial C)} =
a\frac{\partial}{\partial a}L_{\partial C}^{(3)}
\end{equation}

Again, attention is invited to the fact that all explicit dependence of $ W_{I_r}^{(3)}$ on $ a$ enters exclusively through the logarithms. The additional implicit dependence is evident in the relations \cite{G}

$$
(52a)\hspace{2.5in}
R = \frac{12}{a^2}
\hspace{3.0in}
$$
and

$$
(52b)\hspace{2.5in}
K = \frac{1}{3a}cot\theta_4^0
\hspace{2.5in}
$$
as well as

$$
(52c)\hspace{1.5in}
\int_Cd^n\eta = a^42\pi^2(-\frac{1}{3}sin^2\theta_4^0cos\theta_4^0 -
\frac{2}{3}cos\theta_4^0 + \frac{2}{3})
\hspace{1.5in}
$$
and has obviously no consequence upon (46). Moreover, (48) features explicitly both $ L_C^{(3)}(\eta)$ and $ L_{\partial C}^{(3)}(\eta)$ as a result of the commutativity between the operation of double summation over $ N$ and $ N'$ and that of integration over $ C_4$, pursuant to the analysis in Appendices I and II. In effect, a comparison between (48) and (50) reveals that the trace anomaly in the interior of $ C_4$ amounts to

$$
(53)\hspace{1.0in} <T_c^c>_{I_r}^{(C)} =
-\lambda^2RK^2\sum_{N=0}^{\infty}\mathcal{C}_1(N) + \lambda^2R^2
\sum_{N=0}^{\infty}\mathcal{C}_2(N) \hspace{2.0in}
$$
This is the desired local expression for the $ O(\lambda^2)$
contribution to the trace anomaly generated by the renormalised
stress energy-momentum tensor in the interior of $ C_4$. Pursuant
to the results in Appendix II $ \mathcal{C}_1(N)$ and $
\mathcal{C}_2(N)$ are constants for each value of $ N$ and the
infinite series in (53) converge independently to two finite
inconsequential constants. In effect, (53) reveals that, to $
O(\lambda^2)$, the trace anomaly in the interior of $ C_4$
receives contributions exclusively from the $ RK^2$- and $
R^2$-sectors of the vacuum effective action which amount
respectively to the constant values of convergent series over the
quantum number $ N$.

The expression in (53) manifests an exclusive particularity for
the component of this theory's trace anomaly which is generated by
the presence of interactions. Specifically, to $ O(\lambda^2)$ the
trace anomaly in the interior of $ C_4$ receives a constant value.
In  Appendix I the case was also made to the effect that in the
interior, as is also the case in the surface, of $ C_4$ the
effective lagrangian remains respectively constant to any order in
perturbation. Consequently, $ <T_c^c>_{I_r}^{(C)}$ is expected to
receive a constant value in the interior of $ C_4$ to any order.
The origin of the constancy of both the trace anomaly and
effective lagrangian can be traced to the high degree of symmetry
of $ C_4$ which allows for the reduction of the eigenvalue problem
on it to that on $ S_4$ with its characteristic summations over
spherical harmonics of degree $ N$ \cite{G}.

Finally, a comparison between (48) and (51) reveals that

$$
(54)\hspace{1.5in} <T_c^c>_{I_r}^{(\partial C)} =
\lambda^2RK\frac{1}{\Omega_4\pi^4}\frac{1}{sin\theta_4^0}\sum_{N=0}^{\infty}\mathcal{C}_3(N)
\hspace{2.5in}
$$
This is the desired local expression for the $
O(\lambda^2)$ contribution to the trace anomaly on $ \partial
C_4$. It emerges as a contribution from the $ RK$-sector of the
vacuum effective action which amounts to a convergent series over
$ N$ and, as stated, it is also constant.

Both (53) and (54) have emerged as a result of an explicit calculation to $ O(\lambda^2)$. In addition, this calculation reveals the announced result for the contribution to the trace anomaly to $ O(\lambda)$. The finite contribution, associated with the diagram of fig.(1c), is obviously vanishing as that diagram's contribution to the gravitational effective action features no explicit dependence on the embedding radius $ a$ \cite{George}.

In what follows, the renormalisation-group behaviour of the theory will be invoked in order to evaluate the contribution to the trace anomaly to $ O(\lambda^3)$ without resorting to explicit diagramatic calculations.

{\bf V. Trace Anomaly and Renormalisation Group}

In generating perturbatively the vacuum effective action $ W$ the summation over vacuum diagramatic contributions without external wavefunctions also amounts perturbatively to the theory's zero-point proper function. For that matter, the latter coincides identically with $ W$. In effect, (43) is also the $ O(\hbar^3)$ contribution to the theory's zero-point function.

In addition to being generated by the background curvature the divergences contained in $ W_I$ are also the result of the scalar self-interactions. To $ O(\lambda^2)$ they are represented as simple poles at the dimensional limit in (43). The divergences inherent in fig.1(d) and fig.1(e) are the exclusive result of the diagramatic loop-structures respectively and are, consequently, primitive. On the contrary, the divergence inherent in fig.1(c) is the result of the replacement of the bare self-coupling $ \lambda_0$ by $ \lambda $ through (42) and is, for that matter, an overlapping divergence \cite{George}. Consequently, the cancellation of all divergences up to, at least, $ O(\lambda^2)$ necessitates, as stated, the counterterms contained in (42) in addition to the counterterms contained in the gravitational bare couplings. Moreover, (43) reveals that all divergences inherent in $ W_I$ to $ O(\lambda^2)$ are contained as single poles in the $ R^2$, $ RK^2$ and $ RK$ sectors of the bare gravitational action in (3). Therefore, the above-mentioned gravitational bare couplings are the $ \alpha_0$, $ \theta_0$, and $ \zeta_0$ in (3). No additional mass-type counterterms are required. In particular, to $ O(\lambda^2)$ radiative contributions to the two-point function will generate a contribution to $ \xi_0$ in (1) \cite{George}. However, such a contribution does not affect the renormalised zero-point function $ W_{I_r}$ at the stated order. This is the case because any contribution stemming from the $ R\Phi_0^2$ sector of the bare action necessarily enters $ W_{I_r}$ through the contributions which the semi-classical propagator receives from radiative effects to the two-point function. Consequently, contributions generated by that sector to $ O(\lambda^2)$ will necessarily enter $ W_{I_r}$ at higher orders. As an immediate consequence the renormalised zero-point function $ W_{I_r}$ satisfies, to $ O(\lambda^2)$ the renormalisation-group equation

\addtocounter{equation}{3}
\begin{equation}
\big{[}\mu\frac{\partial}{\partial\mu} + \beta(\lambda_r)\frac{\partial}{\partial\lambda_r}
+ \gamma_{\alpha}(\lambda_r)\frac{\partial}{\partial \alpha_r} + \gamma_{\theta}(\lambda_r)\frac{\partial}{\partial \theta_r} + \gamma_{\zeta}(\lambda_r)\frac{\partial}{\partial \zeta_r}\big{]}W_{I_r}^{(3)}(\lambda_r, \alpha_r, \theta_r, \zeta_r) = 0
\end{equation}
with $ \mu$ signifying the usual one-parameter ambiguity of renormalisation and with the renormalisation-group functions being expressed respectively in terms of the renormalised parameters $ \lambda_r, \alpha_r, \theta_r, \zeta_r$ as

$$
(56a)\hspace{2.5in}\beta(\lambda_r) = \mu\frac{\partial\lambda_r}{\partial\mu}\hspace{2.5in}
$$

$$
(56b)\hspace{2.5in}\gamma_{\alpha}(\lambda_r) =  \mu\frac{\partial\alpha_r}{\partial\mu}\hspace{2.5in}
$$

$$
(56c)\hspace{2.5in}\gamma_{\theta}(\lambda_r) =  \mu\frac{\partial\theta_r}{\partial\mu}\hspace{2.5in}
$$

$$
(56d)\hspace{2.5in}\gamma_{\zeta}(\lambda_r) =  \mu\frac{\partial\zeta_r}{\partial\mu}\hspace{2.5in}
$$

Eq.(55) signifies an extension of the general renormalisation-group equation for the vacuum effective action on a curved manifold \cite{Birrel}, \cite{OBS} to the case of a bounded curved manifold. As such, in addition to the $ \beta$- and $ \gamma_{\alpha}$-functions, it features the $ \gamma_{\theta}$-function which accounts for the interaction between volume and surface terms in the interior as well as the $ \gamma_{\zeta}$-function which signifies such an interaction on the boundary.

The stated equation is, essentially, the renormalisation-group equation for a $ n$-point proper function in the special case of $ n=0$. Of course, such an equation becomes perturbatively non-trivial above one-loop level as the one-loop contribution to the zero-point proper function stems exclusively from the free propagation featured in the diagram of fig.1(b). The general importance of the zero-point function to the renormalisation-group behaviour of the theory and evaluation of the trace anomaly stems from the fact that it is entirely independent of the state of the quantum fields. Such an independence is evident in the diagrams of fig.1. As the effective action is the generating functional for proper vertices it follows that $ W$ also coincides  with the theory's zero-point proper function which, in the present case, is perturbatively associated with the vacuum diagrams of fig.1.

As a direct consequence of the scalar vacuum contributions to the $ R^2, RK^2$ and $ RK$ sectors of the bare gravitational action in (3) the component $ W_{I_r}$ of the renormalised vacuum effective action stemming from the scalar self-interactions can be seen on dimensional grounds - at least to $ O(\lambda^2)$ - to, necessarily, depend on $ \lambda_r$, $ a \mu$ and the renormalised gravitational self-couplings $ \alpha_r$, $ \theta_r$ and $ \zeta_r$ only. The above-stated dependence is a consequence of dimensional analysis and is, for that matter, a mathematical property of $ W_{I_r}$. Consequently, the mathematical statement

\addtocounter{equation}{1}
\begin{equation}
dW_{I_r}^{(3)} = \frac{\partial W_{I_r}^{(3)}}{\partial\lambda_r}d\lambda_r +
\frac{\partial W_{I_r}^{(3)}}{\partial (a\mu)}d(a\mu) +
\frac{\partial W_{I_r}^{(3)}}{\partial \alpha_r}d\alpha_r +
\frac{\partial W_{I_r}^{(3)}}{\partial \theta_r}d\theta_r +
\frac{\partial W_{I_r}^{(3)}}{\partial \zeta_r}d\zeta_r
\end{equation}
is distinct from the renormalisation-group statement in (55) which stems from the physical demand that the unrenormalised vacuum effective action be independent of any mass scale. In view of the absence of any mass dimensionality in the renormalised gravitational self-couplings which, again, renders them necessarily functions of $ \lambda$ and $ a\mu$ this statement yields

\begin{equation}
a \frac{\partial}{\partial a}W_{I_r}^{(3)}(\lambda, a\mu) = \mu'\big{[}\frac{\partial W_{I_r}^{(3)}}{\partial\mu'} +
\frac{\partial W_{I_r}^{(3)}}{\partial \alpha_r}\frac{\partial\alpha_r}{\partial\mu'} +
\frac{\partial W_{I_r}^{(3)}}{\partial \theta_r}\frac{\partial\theta_r}{\partial\mu'} +
\frac{\partial W_{I_r}^{(3)}}{\partial \zeta_r}\frac{\partial\zeta_r}{\partial\mu'}\big{]}
\end{equation}
with the redefinition $ \mu'=a\mu$ which does not affect the mass-dimensionality of $ W_{I_r}$ and that of the renormalisation group functions in (56).

In the context of (55) and (56) relation (58) results in

\begin{equation}
a \frac{\partial W_{I_r}}{\partial a} =
-\beta(\lambda_r)\frac{\partial W_{I_r}}{\partial\lambda_r}
\end{equation}
where the $ O(\hbar^3)$-related upper index in $ W_{I_r}^{(3)}$ has been dropped as this procedure  can be repeated in the presence of additional couplings to yield, at any order, the general result in (59).

Eq.(59) signifies a formal relation between the trace anomaly and the renormalisation-group behaviour of the vacuum effective action. In effect, it will provide the basis for the perturbative evaluation of the trace anomaly to $ O(\lambda^3)$, a situation which signifies a loop-order of, at least, four. In its context (46) becomes

\begin{equation}
\int_Cd^4\eta <T^c_c>_{I_r} =
-\beta(\lambda_r)\frac{\partial W_{I_r}}{\partial\lambda_r}
\end{equation}
Again, as was the case in (49), both the integral over $ C_4$ and the vacuum effective action $ W_{I_r}$ in (60) can be resolved respectively into their interior and surface components so that

$$
\int_Cd^4\eta <T_c^c>_{I_r}^{(C)} + \oint_{\partial C}d^3\eta <T_c^c>_{I_r}^{(\partial C)} $$

\begin{equation}
= -\beta(\lambda_r)\frac{\partial}{\partial \lambda_r}\big{[}\int_Cd^4\eta
L_C(\eta) + \oint_{\partial C}d^3\eta L_{\partial C}(\eta)\big{]}
\end{equation}
As stated, both $ L_C(\eta)$ and the interior local component $ <T_c^c>_{I_r}^{(C)}$ receive a constant value at any order in perturbation. Consequently,

$$
<T_c^c>_{I_r}^{(C)}\int_Cd^4\eta + <T_c^c>_{I_r}^{(\partial C)}\oint_{\partial C}d^3\eta = $$

\begin{equation}
-\beta(\lambda_r)\frac{\partial}{\partial \lambda_r}
L_C\int_Cd^4\eta -\beta(\lambda_r)\frac{\partial}{\partial \lambda_r} L_{\partial C}\oint_{\partial C}d^3\eta
\end{equation}
In effect,

\begin{equation}
<T_c^c>_{I_r}^{(C)} = -\beta(\lambda_r)\frac{\partial}{\partial \lambda_r}L_C
\end{equation}
and

\begin{equation}
<T_c^c>_{I_r}^{(\partial C)} = -\beta(\lambda_r)\frac{\partial}{\partial \lambda_r}L_{\partial C}
\end{equation}

The mathematical statements in (63) and (64) are, respectively, the interior and surface aspect of the announced relation between the trace anomaly and the renormalisation-group function for the scalar self-coupling. This relation underscores the relevance of the renormalisation-group behaviour of the theory to its associated conformal anomaly. Attention is invited to the fact that, contrary to the perturbative character of (55), the relations (63) and (64) are, respectively, valid to any order. For that matter they also have formal significance. For instance, any perturbative contribution to the renormalisation-group equation in (55) stemming from $ \xi_0$ will cancel against the corresponding contribution to (57).

In what follows, the perturbative character of (63) and (64) will be exploited in order to advance the evaluation of the trace anomaly to $ O(\lambda^3)$. As the relevant expressions are, as always, particularly involved the calculation will be cited again in some detail.

The beta function $ \beta(\lambda_r)$ of the theory admits the form \cite{Birrel}

$$
(65a)\hspace{2.0in}
\beta(\lambda_r) \equiv
\mu\frac{\partial\lambda_r}{\partial\mu} = (1 - \lambda_r\frac{\partial }{\partial\lambda_r})a_1(\lambda_r)
\hspace{2.0in}
$$
with $ a_1$ being the associated residue of the simple pole in (42) which, to $ O(\lambda^2)$ is \cite{George}

$$
(65b)\hspace{1.5in}
a_1(\lambda_r) = \lambda^2\frac{9}{4\pi^2}(V_c + 1)\sum_{N=1}^{\infty}\frac{\big{[}C_N^{\frac{3}{2}}(cos\theta_4^0)\big{]}^2}{N^2(N + 3)^2}\frac{K^2}{R}
\hspace{1.5in}
$$
Consequently, (65a) yields

\addtocounter{equation}{1}
\begin{equation}
\beta(\lambda_r) = - \lambda^2\frac{9}{4\pi^2}(V_c + 1)\sum_{N=1}^{\infty}\frac{\big{[}C_N^{\frac{3}{2}}(cos\theta_4^0)\big{]}^2}{N^2(N + 3)^2}\frac{K^2}{R}
\end{equation}

It is worth emphasising that the dependence on geometry in (65a)
and (65b) is exclusively specific to $ C_4$. Such a dependence on
a general manifold would contradict the principles of locality and
covariance as a result of which all renormalisation constants must
be space-time independent \cite{OBS}. However, both $ R$ and $ K$
are constants on $ C_4$. There is, for that matter, no
contradiction. Moreover, it is readily seen through (52a) and
(52b) that even the dependence on the embedding radius $ a$
cancels out leaving a trivial dependence on the boundary-defining
angle $ \theta_4^0$.

The renormalised vacuum effective action $ W_{I_r}$ has been
evaluated to $ O(\lambda^2)$ in Appendix II. The result $
W_{I_r}^{(3)}$ has been cited in (48). The derivative of $
L_C^{(3)}$ associated with (63) can be elicited through direct
differentiation with respect to $ \lambda$ of $ W_{I_r}^{(3)}$ in
the interior of $ C_4$ and the replacement of all finite
summations over $ N'$ by the same constants $ \mathcal{C}_1(N)$
and $ \mathcal{C}_2(N)$ as those in (53). Since such
differentiation features all possible $ O(\lambda)$ terms for the
partial derivative of the vacuum effective action with respect to
the scalar self-coupling in the interior of $ C_4$ and (66)
features the $ O(\lambda^2)$ contribution to the beta function it
follows that the complete $ O(\lambda^3)$ contribution to the
trace anomaly in the interior of $ C_4$ is the result of the
substitution in (63) of, both, (66) and the stated partial
derivative. It can readily be confirmed by inspection that such a
contribution also signifies the complete $ O(\hbar^4)$ to the
trace anomaly in the interior of $ C_4$. Moreover, since to $
O(\lambda^2)$ the vacuum effective action $ W_{I_r}^{(3)}$
receives contributions in the $ RK^2$ and $ R^2$ sectors it can be
seen, through (66) and (63), that to $ O(\lambda^3)$ the
contribution to the trace anomaly in the interior of $ C_4$
receives contributions exclusively in the, already established, $
RK^2$-sector as well as in the new $ K^4$-sector which emerges,
for that matter, at loop-orders no lower than four. In effect, to
$ O(\lambda^3)$ to the trace anomaly in the interior of $ C_4$ has
the structure

\begin{equation}
<T^c_c>^{(C)}_{I_r} =
\lambda^3RK^2\sum_{N=0}^{\infty}\mathcal{C}_3(N) -
\lambda^3K^4\sum_{N=0}^{\infty}\mathcal{C}_4(N)
\end{equation}
As was the case to $ O(\lambda^2)$ the contributions to the two
sectors in (67) respectively amount to the constant values of the
associated convergent series over the quantum number $ N$ and can
be read off the results in Appendix II.

As a direct consequence of the above-stated sectors in the trace
anomaly this calculation also reveals in the context of (46) that
to $ O(\lambda^3)$ the renormalised vacuum effective action in the
interior of $ C_4$ develops, itself, a new sector proportionate to
$ K^4$. Equivalently, to the stated order the theory contains a
divergence proportionate to $ K^4$. This is also a qualitatively
new result obtained through the renormalisation-group invariance
of the theory without recourse to explicit diagramatic
calculations.

Finally, the derivative of $ L_{\partial C}^{(3)}$ associated with
(64) can be, likewise, elicited through direct differentiation of
the surface components of $ W_{I_r}^{(3)}$ featured in Appendix
II. Again, the complete $ O(\lambda^3)$ contribution to the trace
anomaly on $ \partial C_4$ is the result of the substitution in
(64) of, both, (66) and the stated partial derivative. It can,
likewise be confirmed by inspection that such a contribution also
signifies the complete $ O(\hbar^4)$ contribution to the trace
anomaly on $ \partial C_4$. As is the case in the interior a new
sector appears in the trace anomaly on $ \partial C$ at $
O(\lambda^3)$. Specifically, the trace anomaly on $ \partial C_4$
receives a contribution proportionate to $ K^3$ at the stated
order which reveals, in effect, that the contribution to the trace
anomaly on $ \partial C_4$ to $ O(\lambda^3)$ has the structure

\begin{equation}
<T^c_c>^{(\partial C)}_{I_r} =
-\lambda^3K^3\sum_{N=0}^{\infty}\mathcal{C}_5(N)
\end{equation}
As a direct consequence the renormalised vacuum effective action
on $ \partial C_4$ develops, itself, a sector proportionate to $
K^3$ at that order in $ \lambda$ or, equivalently, at $
O(\hbar^4)$. To that order this new sector is the result of a
divergence proportionate to $ K^3$ in the theory. Again this is a
qualitatively new result obtained through the
renormalisation-group invariance of the theory.

The results relevant to (67) and (68) are straightforward. For
that matter, it must be emphasised that had the
renormalisation-group approach not been used on the results
obtained from diagramatic evaluations up to second order $
O(\lambda^2)$ in the scalar self-coupling the evaluation of the
trace anomaly as well as that of the vacuum effective action to $
O(\lambda^3)$ through use of (46) would have necessitated explicit
diagramatic calculations of up to - at least - fourth-loop order,
a  rather unenviable task.

{\bf VI. Cosmological Considerations.}

Conformal symmetry breaking is closely associated with the issue
of particle production. As dimensional regularisation is
characterised by the absence of radiative generation of mass the
breaking of any semi-classical conformal invariance, in the
context of that regulating scheme, can only be the result of the
space-time curvature. Such an effect has been explored in the
context of the Friedman-Robertson-Walker space-time \cite{BirDav},
\cite{BO}.

The theory hitherto pursued is predicated on a massless scalar
field conformally coupled to the geometry of $ C_4$ at
semi-classical level. Since, as stated in Sec.III, $ C_4$ is
conformal to the Euclidean, four-dimensional bounded disc $
\mathcal{D}$ there is no particle generation for the stated
massless conformal field at semi-classical level. At one-loop
level stemming from free propagation the presence of a
non-vanishing trace in the quantum expectation value of the stress
tensor, analyzed in Sec.III, has a substantial effect on the
red-shift which the energy density $ \rho$ of the field typically
undergoes. Specifically, an expansion phase for a
three-dimensional spherical hypersurface of equal extrinsic
curvature $ K$ emerges if the colatitude $ \theta_4$ which
characterises $ \partial C$ with $ K > 0$ is associated with the
imaginary time parameter $ \tau$ through the relation $ \tau =
a\theta$, with $ a$ being the embedding radius. In effect, such an
approach renders the stated hypersurface of equal extrinsic
curvature also as one of equal time.

As a result of such an expansion the energy density of the
conformal field at semi-classical level undergoes a red-shift from
some initial value $ \rho(\tau_1)$ to the final value $
\rho(\tau_2)$. The breaking of conformal symmetry from free
propagation which occurs at one-loop level alters the stated
energy density's red-shift. If $ \rho(0)$ corresponds to the
energy density of a vanishing three geometry and $
\rho(\tau_{\partial C})$ to that corresponding to the boundary
hypersurface then the emergence of the anomalous trace of the
stress tensor relates those two values as \cite{BirDav}

\begin{equation}
\rho(\tau_{\partial C})\Omega^4(\tau_{\partial C}) =
\rho(0)\Omega^4(0) - \int_{0}^{\tau_{\partial
C}}d\tau\Omega^3(\tau)\frac{d}{d\tau}\Omega(\tau)<T^c_c>_r^{(C)}
\end{equation}
with $ \Omega(\tau)$ being the conformal factor relating the
geometry of $ C_4$ to that of $ \mathcal{D}$.

The relation expressed by (69) is, in fact, relevant to all
Robertson-Walker space-times. The four-dimensional Euclidean de
Sitter space, a bounded segment of which is $ C_4$, is a
particular case of a Robertson-Walker space-time, albeit in
imaginary time. In Robertson-Walker space-times the integrand in
the second term of the right side in (69) amounts to a pure time
derivative in the case of free propagation. For that matter the
integral in (69) vanishes if the boundaries of that integration
correspond to asymptotic regions in which the space-time is static
\cite{BirDav}. As this is, obviously, not the case on $ C_4$ it
would appear that the stated expansion of the three-dimensional
spherical sections results in particle generation. However, such a
generation need not emerge from the absence of such static
asymptotic regions on $ C_4$. In fact, if the scalar field is in a
conformal vacuum state at any specific time then it shall remain
in that state at all times. Any particle generation detected by a
co-moving observer in any space-time conformal to Minkowski
space-time in real time, or to flat Euclidean space in imaginary
time, merely reveals the irrelevance of the intuitive particle
concept in curved space-times \cite{Birrel}. Consequently, even if
the relevant manifold does not have static, asymptotic regions the
relation in (69) may still be interpreted as a red-shift for the
energy density on condition that the field is in a conformal
vacuum state.

The bounded manifold $ C_4$, however, is conformal only to a
bounded segment $ \mathcal{D}$ of the Euclidean four-dimensional
space. For that matter, no meaningful definition of any vacuum
state of $ C_4$ as conformal vacuum can be given. Indeed, as a
direct consequence of the absence of Poincare invariance all
observers on $ \mathcal{D}$ regardless of their state of motion
will detect gravitons associated with the curvature of $ \partial
\mathcal{D}$. As a result, even for an observer in free fall is
there no state on $ C_4$ characterised by the absence of massless
scalar particles nor, for that matter, by the absence of
gravitons. Although such quanta have no contribution to $
<T^c_c>_r^{(C)}$ they result in a non-vanishing value for the
integral in (69). In turn, additional particles are generated by
the expansion through the anomalous trace.

In addition to the stated process yet another source of particle
generation can be elicited by (32) and (39). Specifically,
although the integrand in (69) is a pure time derivative in the
interior of $ C_4$ it receives a contribution from the anomalous
trace on $ \partial C$ which does not amount to a time derivative.
Such a contribution amounts directly to a contribution in the
conformal field's energy density generated by the anomalous trace
in the course of the expansion. Contrary to the situation in
Robertson-Walker type manifolds conformal symmetry breaking
emerging from free propagation results in particle generation on $
C_4$ through the two above stated sources.

The process of particle generation described above in the context
of free propagation is substantially enhanced if $
<T^c_c>_r^{(C)}$ contains terms which do not in general, lead to
pure time derivatives in (69). This is the case in the presence of
vacuum effects generated by mutually interacting matter fields.
Specifically, terms in the anomalous trace which are of the form

$$
g^{-\frac{1}{2}}(x)\beta(\lambda_r)\frac{\delta
W_{I_r}(\lambda_r)}{\delta\lambda_r}
$$
result in particle generation although, classically, such
generation is precluded \cite{BirDav}. This expression reveals
that in the context of the present theory such terms arise
directly from the renormalisation of the scalar self- coupling $
\lambda_0$ in (1) and are perturbatively contained in (60). The
effect of this contribution on the final energy density can be
obtained from (69). As was also the situation in the case of free
propagation the stated contribution can be seen, through (63) and
(64), to contain an interior as well as a boundary component.


The generation of particles as a result of conformal symmetry
breaking is an example of how a long range effect can emerge as a
result of vacuum processes \footnote{The conformal anomaly as a
measure of particle production at infinity in the context of the
Hawking radiation is another such example.}. In effect, such
particle generation is relevant to the quantisation of closed
cosmological models and the no-boundary proposal \cite{HarHawk}.

As stated in Sec.III if, in the case of $ \Lambda > 0$, a boundary
condition imposed on the Euclidean Einstein field equations
relevant to a massless scalar field conformally coupled to gravity
is that of a compact four-geometry then the solution is $ C_4$ on
the additional Dirichlet condition of a constant field on $
\partial C_4$. In effect, if only compact geometries and regular
fields are considered then at the semi-classical limit $ C_4$ is
the geometry with the dominant contribution to the Euclidean path
integral which yields the probability amplitude for the
three-geometry of $ \partial C_4$ and constant scalar field
specified on it. The results obtained herein regarding the
perturbative evaluation of the trace anomaly as well as the
perturbative evaluation of the effective action and the
renormalisation of the theory on $ C_4$ remain intact if $ K<0$ on
$ \partial C_4$. For that matter, they also characterise the
cosmological semi-classical tunnelling geometry which consists in
a Riemannian segment joined smoothly onto half the Euclidean
sphere at the space-like spherical section of $ K=0$. The exact
expressions for the associated results on that manifold will
emerge through an appropriate analytical extension on the
Euclidean-time coordinate inherent in $ C_4$ as well as on the
negative values of the Euclidean $ K$. The stated semi-classical
geometry accommodating the conformal scalar field will provide the
dominant contribution to the probability amplitude for the
boundary space-like three-sphere and constant scalar field thereon
in real time.

The length scale relevant to the tunnelling process is the
Euclidean sphere's embedding radius $ a$ which is, itself,
inversely proportional to the square root of the positive
cosmological constant. For that matter, in the context of such a
process the conformal anomaly, perturbatively evaluated herein, is
expected to result in the stated long range effect of particle
production even in cosmological models of very small $ \Lambda$.
Although the result of vacuum processes described by the
contributions of the relevant diagrams in fig.1 the massless
quanta will still be present at the semi-classical scale
associated with the expanding boundary space-like three-sphere of
the tunnelling geometry. Such an effect naturally has an impact on
the probability for tunnelling. Although conformal invariance is
upheld at semi-classical level the presence of those massless
quanta potentially modifies the stress tensor and, for that
matter, affects the tunnelling geometry itself. Although such an
effect is small if the cosmological constant itself is small it is
of interest if only because it is not vanishing.

{\bf VII. Concluding Remarks.}

The conformal anomaly for a massless self-interacting scalar field
conformally coupled to the background geometry of a bounded
manifold has been pursued and evaluated in the context of free
scalar propagation as well as in the higher-loop context generated
by the presence of scalar self-interactions to second and third
order in the scalar self-coupling. In the free case, the results
in the interior and boundary of the manifold are, respectively,
cited in expressions (38) and (41). In the context of the
perturbative evaluation of the effective action in Appendix II to
second order in the scalar self-coupling the corresponding results
in the interior and boundary are, respectively, cited in
expressions (53) and (54). The corresponding results in the
interior and boundary to third order in the scalar self-coupling
are, respectively, cited in (67) and (68). The contribution to the
trace anomaly stemming at order one in the scalar self-coupling is
vanishing. The evaluation of up to order three in the scalar
self-coupling, accomplished herein, involves the complete
contribution which the conformal anomaly receives from the fourth
order in the loop expansion.

For reasons of technical convenience the stated evaluation has
been pursued on a Riemannian manifold of positive constant
curvature accommodating a boundary of positive extrinsic curvature
with homogeneous Dirichlet-type boundary conditions.
Notwithstanding convenience, however, the results obtained herein
regarding the perturbative evaluation of the trace anomaly as well
as the perturbative evaluation of the effective action and the
renormalisation of the theory on $ C_4$ remain intact if $ K<0$ on
$ \partial C_4$. For that matter, they also characterise the
cosmological semi-classical tunnelling geometry which consists in
a Riemannian segment joined smoothly onto half the Euclidean
sphere at the space-like spherical section of $ K=0$. The exact
expressions for the associated results on that manifold will
emerge through an appropriate analytical extension on the
Euclidean-time coordinate inherent in $ C_4$ as well as on the
negative values of the Euclidean $ K$. In the context of the
no-boundary proposal \cite{HarHawk} the stated semi-classical
geometry accommodating the conformal scalar field will provide the
dominant contribution to the probability amplitude for the
boundary space-like three-sphere and constant scalar field thereon
in real time.

The evaluation of the theory's trace anomaly reveals information about
the theory's vacuum effective action to the same order. In addition to
the $ RK^2$ and $ R^2$-sectors established to $ O(\lambda^2)$ in the
interior of $ C_4$ vacuum effects generate a $ K^4$-sector in the
renormalised vacuum effective action to $ O(\lambda^3)$ or, equivalently,
to fourth loop-order. Likewise, in addition to the established $ RK$-sector,
the renormalised vacuum effective action develops a $ K^3$-sector on
$ \partial C_4$ to the same order. Dimensionally, these are the only new
sectors which the vacuum effective action can accommodate.

These results directly allow for a perturbative evaluation of the vacuum
effective action to fourth loop-order. The renormalised vacuum effective
action can be obtained from the conformal anomaly if the latter has already
been, independently, attained \cite{Riegert}. In the present theory that is,
perturbatively, the case to $ O(\lambda^3)$. For that matter, the exact form
of the renormalised gravitational effective action at four-loop order
$ W_{I_r}^{(4)}$, which is also the entire contribution to $ O(\lambda^3)$,
can be directly elicited from (66), (63) and (64) as well as Appendix II-related (48).

In addition to the immediate importance which the perturbative
evaluation of the theory's trace anomaly and effective action at
four-loop order has for the dynamical behaviour of quantised
fields on bounded manifolds the results herein are also of
importance in quantum cosmology.

{\bf Acknowledgements}

I would like to thank Sergei Odintsov for bringing a relevant reference to my attention.

{\bf Appendix}

{\bf I. The ``massless tadpole" revisited}

The massless scalar propagator on $ C_n$ is \cite{G}

$$ D_C^{(n)}(\eta, \eta') = \frac{1}{|\eta - \eta'|^{n-2}} - \frac{1}{|\frac{a_{\eta'}}{a_B}\eta - \frac{a_B}{a_{\eta'}}\eta'|^{n-2}}
$$
and features explicitly the geodesic distance $ a_B$ between the pole and the boundary $ \partial C_n$ as well as the geodesic distance $ a_{\eta'}$ between the pole and point $ \eta'$ on $ C_n$. In the context of the method of images both the singular and boundary part of the propagator on $ C_n$ are independently treated as propagators on $ S_n$. The corresponding expansions in terms of spherical harmonics in transform space are respectively \cite{G}

$$
|\eta - \eta'|^{2-n} = \sum_{N=0}^{\infty}\sum_{\alpha = 0}^N\big{[}- \frac{a^2}{(N + \frac{n}{2} - 1)(N + \frac{n}{2})}\big{]}Y_{\alpha}^N(\eta)Y_{\alpha}^N(\eta')
$$
and likewise

$$
|\frac{a_{\eta'}}{a_B}\eta - \frac{a_B}{a_{\eta'}}\eta'|^{2-n} = \sum_{N=0}^{N_0}\sum_{\alpha = 0}^N\big{[}- \frac{a^2}{(N + \frac{n}{2} - 1)(N + \frac{n}{2})}\big{]}Y_{\alpha}^N(\frac{a_{\eta'}}{a_B}\eta)Y_{\alpha}^N(\frac{a_B}{a_{\eta'}}\eta')
$$
The singular part of $ D_C^{(n)}(\eta, \eta')$ contains the
singularity at the coincidence limit $ \eta \rightarrow \eta'$ on
$ C_n$. The exact nature of $ N_0$ has been analyzed in \cite{G}.
In summary, the absence of a coincidence limit for
image-propagation on $ S_n$ stemming from the boundary part of the
propagator signifies a cut-off separation below which
image-propagation on $ S_n$ vanishes. In the context of
image-propagation this, in turn, results in the finite upper limit
$ N_0$ for the quantum number $ N$ which in transform space is
associated with the angular momentum flowing through the boundary
part of the propagator \cite{G}.

The expression for the one-loop diagramatic contribution to the two-point Green function for a massless conformal scalar field on $ C_4$ has been derived in \cite{G} and reproduced in \cite{George}. The ensuing analysis is best advanced if the proper diagram rather than the connected one is considered. The result for the proper diagram is

$$
(A1)\hspace{0.5in}
(-\lambda)\int_{C_4}d^4\eta D^{(4)}(\eta, \eta) = -(-\lambda)\frac{1}{3}\frac{1}{2^{10}}\frac{V_c + 1}{\pi^{\frac{7}{2}}}\big{[}\sum_{N=0}^{N_0}(2N +
3)\big{]}R\int_{C_4}d^4\eta
\hspace{0.5in}
$$
and emerges at the dimensional limit $ n \rightarrow 4$ on the expression

$$
(A2)\hspace{0.5in}
(-\lambda)\int_{C_n}d^n\eta D^{(n)}(\eta, \eta) =
-(-\lambda)\frac{1}{8\pi^{\frac{n}{2}}}\Gamma(\frac{n - 1}{2})\frac{1}{(n - 2)!}a^{2-n}
\times
\hspace{1.5in}
$$

$$
\sum_{N=0}^{N_0}\frac{(2N + n - 1)\Gamma(N + n - 1)}{(N + \frac{n}{2})(N + \frac{n}{2} - 1)\Gamma(N + 1)}\int_{C_n}d^n\eta
$$
where the symmetry factor of $ \frac{1}{2}$ has been included.

At the coincidence limit $ \eta \rightarrow \eta'$ the singular part of the propagator $ D^{(n)}(\eta, \eta')$ integrated on $ C_n$ in $ (A2)$ vanishes. At the limit $ \eta \rightarrow \eta'$ the boundary part of the propagator $ D^{(n)}(\eta, \eta')$ integrated on $ C_n$ is responsible for the result cited in $ (A2)$ through

$$
(A3)\hspace{0.5in}
(-\lambda)\int_{C_n}d^n\eta D^{(n)}(\eta, \eta) =
-(-\lambda)\frac{1}{8\pi^{\frac{n}{2}}}\Gamma(\frac{n - 1}{2})\frac{1}{(n - 2)!}a^{2-n}
\times
\hspace{1.5in}
$$

$$
\int_{C_n}d^n\eta\sum_{N=0}^{N_0}\frac{(2N + n - 1)\Gamma(N + n - 1)}{(N + \frac{n}{2})(N + \frac{n}{2} - 1)\Gamma(N + 1)}
$$

Both $ (A2)$ and $ (A3)$ feature the transform-space cut-off $ N_0$ for image propagation on $ S_n$. However, $ N_0$ is contingent upon the point of $ C_n$ at which the coincidence limit $ \eta \rightarrow \eta'$ occurs prior to integration. This is immediately obvious from the fact that propagation on $ C_n$ vanishes if either $ \eta$ or $ \eta'$ in $ D^{(n)}(\eta, \eta')$ is on $ \partial C_n$ as a result of which situation $ N_0 \rightarrow \infty$ \cite{G}. In fact, on account of the symmetry which characterises $ C_n$ the cut-off $ N_0$ increases as the geodesic distance between  $ \partial C_n$ and the equal imaginary-time spherical hypersurface of equal extrinsic curvature on which the  limit $ \eta \rightarrow \eta'$ is taken prior to integration on $ C_n$ decreases. For that matter, the commutativity between the operation of summation over $ N$ and that of integration over $ C_n$ is  non-trivial and a careful analysis is necessary in order to derive $ (A2)$ from $ (A3)$. As the objective in \cite{G} was the exploration of the general qualitative aspects of the theory this analysis was not explicitly advanced.

The fact that the integer $ N_0$ increases indefinitely as the hypersurface of the coincidence limit approaches $ \partial C_n$ implies that the interior of $ C_n$ is segregated by a set of hypersurfaces each of which is defined at a fixed geodesic distance from $ \partial C_n$ and is, consequently, characterised by constant extrinsic curvature and constant imaginary time. On each segment defined by two such successive hypersurfaces
$ N_0$ receives a constant integer value. That value increases to $ N_0+1$ on the next segment characterised by a smaller volume and a smaller geodesic separation between its biggest hypersurface and $ \partial C_n$. The fact that $ N_0 \rightarrow \infty$ as the geodesic distance between the biggest hypersurface of a segment and $ \partial C_n$ tends to zero implies that the density of such segments on $ C_n$ increases, itself, indefinitely toward $ \partial C_n$. In the context of such a segregation the summation over $ N$ independently commutes with integration exclusively over that segment of $ C_n$ which is associated with the relevant $ N_0$. Consequently,

$$
\hspace{1.5in}
\int_{C_n}d^n\eta\sum_{N=0}^{N_0}\frac{(2N + n - 1)\Gamma(N + n - 1)}{(N + \frac{n}{2})(N + \frac{n}{2} - 1)\Gamma(N + 1)} =
\hspace{1.5in}
$$

$$
\int_{0}d^n\eta\sum_{N=0}^{N_0}\frac{(2N + n - 1)\Gamma(N + n - 1)}{(N + \frac{n}{2})(N + \frac{n}{2} - 1)\Gamma(N + 1)} + \int_{1}d^n\eta\sum_{N=0}^{N_0}\frac{(2N + n - 1)\Gamma(N + n - 1)}{(N + \frac{n}{2})(N + \frac{n}{2} - 1)\Gamma(N + 1)} + ...
$$

$$
+ \int_{N_0-1}d^n\eta\sum_{N=0}^{N_0}\frac{(2N + n - 1)\Gamma(N + n - 1)}{(N + \frac{n}{2})(N + \frac{n}{2} - 1)\Gamma(N + 1)} + \int_{N_0}d^n\eta\sum_{N=0}^{N_0}\frac{(2N + n - 1)\Gamma(N + n - 1)}{(N + \frac{n}{2})(N + \frac{n}{2} - 1)\Gamma(N + 1)} + ... =
$$

$$
\frac{(n - 1)\Gamma(n - 1)}{\frac{n}{2}(\frac{n}{2} - 1)}\int_{0}d^n\eta +
\big{[}\frac{(n - 1)\Gamma(n - 1)}{\frac{n}{2}(\frac{n}{2} - 1)} +
\frac{(1 + n)\Gamma(n)}{(1 + \frac{n}{2})\frac{n}{2}}\big{]}\int_{1}d^n\eta + ... +
$$

$$
\sum_{N=0}^{N_0-1}\frac{(2N + n - 1)\Gamma(N + n - 1)}{(N + \frac{n}{2})(N + \frac{n}{2} - 1)\Gamma(N + 1)}\int_{N_0-1}d^n\eta +
\sum_{N=0}^{N_0}\frac{(2N + n - 1)\Gamma(N + n - 1)}{(N + \frac{n}{2})(N + \frac{n}{2} - 1)\Gamma(N + 1)}\int_{N_0}d^n\eta
$$

$$
(A4)\hspace{1.5in}
+\sum_{N=0}^{N_0+1}\frac{(2N + n - 1)\Gamma(N + n - 1)}{(N + \frac{n}{2})(N + \frac{n}{2} - 1)\Gamma(N + 1)}\int_{N_0+1}d^n\eta + ...
\hspace{1.5in}
$$

with

$$
(A5)\hspace{2.5in}
\sum_{N=0}^{\infty}\int_{N}d^n\eta = \int_{C_n}d^n\eta
\hspace{2.5in}
$$
However, it is \cite{G}

$$
(A6)\hspace{1.5in}
\sum_{N=0}^{\infty}\frac{(2N + n - 1)\Gamma(N + n - 1)}{(N + \frac{n}{2})(N + \frac{n}{2} - 1)\Gamma(N + 1)} = 0
\hspace{1.5in}
$$
This is the case because for $ n<2$ the only negative infra-red $ (N=0)$-term cancels
identically against the ultra-violet sum-total of the remaining infinite number of
positive terms. (A6) subsequently emerges as the result of an analytical extension at
$ n=4$ \cite{Drummond}. For that matter,

$$
\big{|}\sum_{N=0}^{N_0+1}\frac{(2N + n - 1)\Gamma(N + n - 1)}{(N + \frac{n}{2})(N + \frac{n}{2} - 1)\Gamma(N + 1)}\big{|} < \big{|}\sum_{N=0}^{N_0}\frac{(2N + n - 1)\Gamma(N + n - 1)}{(N + \frac{n}{2})(N + \frac{n}{2} - 1)\Gamma(N + 1)}\big{|}
$$
and consequently

$$
\sum_{N=0}^{N_0+1}\frac{(2N + n - 1)\Gamma(N + n - 1)}{(N + \frac{n}{2})(N + \frac{n}{2} - 1)\Gamma(N + 1)}\int_{N_0+1}d^n\eta =
\sum_{N=0}^{N_0}\frac{(2N + n - 1)\Gamma(N + n - 1)}{(N + \frac{n}{2})(N + \frac{n}{2} - 1)\Gamma(N + 1)}\int_{<(N_0+1)}d^n\eta
$$
with the integral associated with the notation $ <(N_0+1)$ corresponding to the volume of an appropriate portion of that segment on $ C_n$ on which the transfer-space cut-off on $ S_n$ is $ N_0+1$ and for which integral, in effect, it is

$$
\int_{<(N_0+1)}d^n\eta < \int_{N_0+1}d^n\eta
$$
This allows for

$$
\sum_{N=0}^{N_0}\frac{(2N + n - 1)\Gamma(N + n - 1)}{(N + \frac{n}{2})(N + \frac{n}{2} - 1)\Gamma(N + 1)}\int_{N_0}d^n\eta + \sum_{N=0}^{N_0+1}\frac{(2N + n - 1)\Gamma(N + n - 1)}{(N + \frac{n}{2})(N + \frac{n}{2} - 1)\Gamma(N + 1)}\int_{N_0+1}d^n\eta =
$$

$$
\sum_{N=0}^{N_0}\frac{(2N + n - 1)\Gamma(N + n - 1)}{(N + \frac{n}{2})(N + \frac{n}{2} - 1)\Gamma(N + 1)}\big{[}\int_{N_0}d^n\eta + \int_{<(N_0+1)}d^n\eta\big{]}
$$
which, through

$$
\int_{<(N_0+1)}d^n\eta = a_{N_0+1}\int_{N_0}d^n\eta ~~~;~~~0 < a_{N_0+1} < 1
$$
results in

$$
\sum_{N=0}^{N_0}\frac{(2N + n - 1)\Gamma(N + n - 1)}{(N + \frac{n}{2})(N + \frac{n}{2} - 1)\Gamma(N + 1)}\int_{N_0}d^n\eta + \sum_{N=0}^{N_0+1}\frac{(2N + n - 1)\Gamma(N + n - 1)}{(N + \frac{n}{2})(N + \frac{n}{2} - 1)\Gamma(N + 1)}\int_{N_0+1}d^n\eta =
$$

$$
(A7)\hspace{1.5in}
\sum_{N=0}^{N_0}\frac{(2N + n - 1)\Gamma(N + n - 1)}{(N + \frac{n}{2})(N + \frac{n}{2} - 1)\Gamma(N + 1)}\big{[}1 + a_{N_0+1}\big{]}\int_{N_0}d^n\eta
\hspace{2.5in}
$$
where, as will become obvious in what follows, the range $ 0<a_{N_0+1}<1$ is guarranted by
the finite result to which the integral on the left side of (A4) amounts. Substituting (A7) in (A4) results in

$$
\hspace{1.5in}
\int_{C_n}d^n\eta\sum_{N=0}^{N_0}\frac{(2N + n - 1)\Gamma(N + n - 1)}{(N + \frac{n}{2})(N + \frac{n}{2} - 1)\Gamma(N + 1)} =
\hspace{1.5in}
$$

$$
\frac{(n - 1)\Gamma(n - 1)}{\frac{n}{2}(\frac{n}{2} - 1)}\int_{0}d^n\eta +
\big{[}\frac{(n - 1)\Gamma(n - 1)}{\frac{n}{2}(\frac{n}{2} - 1)} +
\frac{(1 + n)\Gamma(n)}{(1 + \frac{n}{2})\frac{n}{2}}\big{]}\int_{1}d^n\eta + ... +
$$

$$
\sum_{N=0}^{N_0-1}\frac{(2N + n - 1)\Gamma(N + n - 1)}{(N + \frac{n}{2})(N + \frac{n}{2} - 1)\Gamma(N + 1)}\int_{N_0-1}d^n\eta +
$$

$$
(A8)\hspace{1.0in}
\sum_{N=0}^{N_0}\frac{(2N + n - 1)\Gamma(N + n - 1)}{(N + \frac{n}{2})(N + \frac{n}{2} - 1)\Gamma(N + 1)}\big{[}1 + a_{N_0+1}\big{]}\int_{N_0}d^n\eta + ...
\hspace{2.0in}
$$

This reduction in (A4) of the first $ N_0+1$ terms of the series in (A6) to the first $ N_0$ terms can be repeated for the first $ N_0+k$ terms in (A6) for an arbitrary value of $ k$. For that matter (A8) is replaced by

$$
\hspace{1.5in}
\int_{C_n}d^n\eta\sum_{N=0}^{N_0}\frac{(2N + n - 1)\Gamma(N + n - 1)}{(N + \frac{n}{2})(N + \frac{n}{2} - 1)\Gamma(N + 1)} =
\hspace{1.5in}
$$

$$
\frac{(n - 1)\Gamma(n - 1)}{\frac{n}{2}(\frac{n}{2} - 1)}\int_{0}d^n\eta +
\big{[}\frac{(n - 1)\Gamma(n - 1)}{\frac{n}{2}(\frac{n}{2} - 1)} +
\frac{(1 + n)\Gamma(n)}{(1 + \frac{n}{2})\frac{n}{2}}\big{]}\int_{1}d^n\eta + ... +
$$

$$
\sum_{N=0}^{N_0-1}\frac{(2N + n - 1)\Gamma(N + n - 1)}{(N + \frac{n}{2})(N + \frac{n}{2} - 1)\Gamma(N + 1)}\int_{N_0-1}d^n\eta +
$$

$$
(A9)\hspace{0.5in}
\sum_{N=0}^{N_0}\frac{(2N + n - 1)\Gamma(N + n - 1)}{(N + \frac{n}{2})(N + \frac{n}{2} - 1)\Gamma(N + 1)}\big{[}1 + a_{N_0+1} + a_{N_0+2} + ...+ a_{N_0+k} + ...\big{]}c_{N_0}\int_{C_n}d^n\eta
\hspace{2.0in}
$$
where, again, the volume of the segment of $ C_n$ featuring the cut-off $ N_0$ has been expressed in terms of the manifold's entire volume through

$$
\int_{N_0}d^n\eta = c_{N_0}\int_{C_n}d^n\eta ~~;~~0 < c_{N_0} < 1
$$

The result in (A9) reveals essentially a choice of an integer value for $ N_0$ prior to integration over $ C_n$ on the left side of the equality. This choice is realised in transform space and reflects the choice of a point $ \eta$ for the coincidence limit on $ C_n$ prior to integration over $ C_n$.

In order to arrive at the desired result cited in (A2) the remaining $ N_0$ summations in (A9) featuring, respectively, values for the transfer-space cut-off not exceeding $ N_0-1$ will also have to be expressed in terms of the summation of the first $ N_0$ terms. Again, this objective shall be realised through

$$
\big{|}\sum_{N=0}^{N_0-1}\frac{(2N + n - 1)\Gamma(N + n - 1)}{(N + \frac{n}{2})(N + \frac{n}{2} - 1)\Gamma(N + 1)}\big{|} > \big{|}\sum_{N=0}^{N_0}\frac{(2N + n - 1)\Gamma(N + n - 1)}{(N + \frac{n}{2})(N + \frac{n}{2} - 1)\Gamma(N + 1)}\big{|}
$$
which implies

$$
\sum_{N=0}^{N_0-1}\frac{(2N + n - 1)\Gamma(N + n - 1)}{(N + \frac{n}{2})(N + \frac{n}{2} - 1)\Gamma(N + 1)}\int_{N_0-1}d^n\eta = \sum_{N=0}^{N_0}\frac{(2N + n - 1)\Gamma(N + n - 1)}{(N + \frac{n}{2})(N + \frac{n}{2} - 1)\Gamma(N + 1)}\int_{>(N_0-1)}d^n\eta
$$
with the integral associated with the notation $ >(N_0-1)$ corresponding to a volume in excess of that of the segment of $ C_n$ on which the transfer-space cut-off on $ S_n$ is $ N_0-1$. As the density of segregation on $ C_n$ increases toward $ \partial C_n$ it is

$$
\int_{>(N_0-1)}d^n\eta = a'_{N_0-1}\int_{N_0}d^n\eta ~~;~~ a'_{N_0-1} > 1
$$
so that repeating this procedure for all stated remaining summations yields

$$
\hspace{1.5in}
\int_{C_n}d^n\eta\sum_{N=0}^{N_0}\frac{(2N + n - 1)\Gamma(N + n - 1)}{(N + \frac{n}{2})(N + \frac{n}{2} - 1)\Gamma(N + 1)} =
\hspace{1.5in}
$$

$$
\sum_{N=0}^{N_0}\frac{(2N + n - 1)\Gamma(N + n - 1)}{(N + \frac{n}{2})(N + \frac{n}{2} - 1)\Gamma(N + 1)}\big{[}a'_{1} + ... + a'_{N_0-2} + a'_{N_0-1}\big{]}c_{N_0}\int_{C_n}d^n\eta +
$$

$$
\hspace{0.5in}
\sum_{N=0}^{N_0}\frac{(2N + n - 1)\Gamma(N + n - 1)}{(N + \frac{n}{2})(N + \frac{n}{2} - 1)\Gamma(N + 1)}\big{[}1 + a_{N_0+1} + a_{N_0+2} + ...+ a_{N_0+k} + ...\big{]}c_{N_0}\int_{C_n}d^n\eta
\hspace{2.0in}
$$
and, finally

$$
\hspace{0.5in}
\int_{C_n}d^n\eta\sum_{N=0}^{N_0}\frac{(2N + n - 1)\Gamma(N + n - 1)}{(N + \frac{n}{2})(N + \frac{n}{2} - 1)\Gamma(N + 1)} =
$$

$$
(A10)\hspace{1.5in}
\Lambda^{(n)}(N_0)\sum_{N=0}^{N_0}\frac{(2N + n - 1)\Gamma(N + n - 1)}{(N + \frac{n}{2})(N + \frac{n}{2} - 1)\Gamma(N + 1)}\int_{C_n}d^n\eta
\hspace{2.0in}
$$
where the $ N_0$-dependent multiplicative factor in n dimensions is defined through

$$
\Lambda^{(n)}(N_0) = \big{[}a'_{1} + ... + a'_{N_0-2} + a'_{N_0-1} + 1 + a_{N_0+1} + a_{N_0+2} + ...+ a_{N_0+k} + ...\big{]}c_{N_0}
$$
and is finite as a result of (A6) or, equivalently, as a result of

$$
0 < a_{N_0+k} < 1 , k = 0, 1, 2, ... ~~;~~ 0 < c_{N_0} < 1
$$
(A10) is the desired result. Replacing (A10) in (A3) results in (A2) with the additional presence of the multiplicative factor $ \Lambda^{(n)}(N_0)$. The final result, for that matter, is

$$
(A11)\hspace{0.5in}
(-\lambda)\int_{C_n}d^n\eta D^{(n)}(\eta, \eta) =
-(-\lambda)\frac{1}{8\pi^{\frac{n}{2}}}\Gamma(\frac{n - 1}{2})\frac{1}{(n - 2)!}a^{2-n}
\times
\hspace{1.5in}
$$

$$
\Lambda^{(n)}(N_0)
\sum_{N=0}^{N_0}\frac{(2N + n - 1)\Gamma(N + n - 1)}{(N + \frac{n}{2})(N + \frac{n}{2} - 1)\Gamma(N + 1)}\int_{C_n}d^n\eta
$$
As the $ N_0$-dependence of the sum on the left side of (A10) has essentially been integrated out over $ C_4$ the cut-off $ N_0$, itself, is intrinsically unspecified on the left side of (A10). For that matter it must be just as  unspecified on the right side of it. Consequently, any subjective choice of $ N_0$ on the right side of (A11), corresponding to the choice of the coincidence limit $ \eta' \rightarrow \eta$ in $ D(\eta, \eta')$, must be offset by $ \Lambda^{(n)}(N_0)$. For that matter, the result for the one-loop diagramatic contribution to the theory's two-point function is independent of $ N_0$. Moreover, as the product between $ \Lambda^{(n)}(N_0)$ and its corresponding sum over $ N_0$ amounts to a $ N_0$-independent constant the exact form of the former is inconsequential to the renormalised effective action and to the renormalisation of the theory itself.

The entire analysis eventuating in the commutativity between the
summation over $ N$ and the integration over $ C_4$ remains valid
for all other summations terminating at $ N_0$ in the context of
the general formulation for diagramatic evaluations on $ C_4$
\cite{G}-\cite{George}. Such summations and their associated
integrals are featured below in Appendix II. All diagramatic
contributions to the effective action stemming from scalar
interactions explicitly manifest the particularity of that
commutativity in their finite summations. As a consequence, the
renormalised effective lagrangian is constant on $ C_4$. This has
been explicitly confirmed to $ O(\lambda^2)$ and will be assumed
to remain valid to all orders in perturbation on the grounds that
the finite summations are the only source of local dependence
through their upper limit $ N_0$. Such a dependence is always
cancelled by $ \Lambda^{(n)}(N_0)$. The constancy of the effective
lagrangian is a particularity of the theory which can be traced to
the reduction of the eigenvalue problem on $ C_4$ to that on $
S_4$.

{\bf Appendix II.}

{\bf $ O(\lambda^2)$-renormalised vacuum effective action.}

In what follows, attention is invited to the transform-space
cut-off $ N_0$ for image propagation on $ C_4$ emerging from the
boundary part of the propagator \cite{G}. As stated, $ N$ is the
quantum number associated with the angular momentum flowing
through the singular part of the propagator and $ N'$ is the
corresponding quantum number associated with the boundary part of
the propagator. The finite sums featuring $ N_0$ have been shown
to emerge from the one-loop diagramatic contribution to the
two-point Green function as well as from the general structure of
the entire formulation for diagramatic evaluations on $ C_4$
\cite{G}, \cite{T}, \cite{GT}, \cite{George}. Again, the cut-off $
N_0$ is contingent upon the geodesic separation between the
coincidence limit $ \eta \rightarrow \eta'$ for the scalar
propagator $ D^{(4)}(\eta, \eta')$ and $ \partial C_4$. The
procedure in Appendix I applied to integrals of the type \cite{GT}

$$
\int_Cd^n\eta d^n\eta'\sum_{N=0}^{\infty}G(N)\sum_{N'=0}^{N'_0}J(N')Y_{\alpha}^N(\eta)Y_{\alpha'}^{N'}(\eta)Y_{\alpha}^N(\eta')Y_{\alpha'}^{N'}(\eta') $$
results in the commutativity between the operation of summation over $ N'$ up to $ N'_0$ and that of integration over $ C_4$ through the finite multiplicative factor $ L^{(n)}(N_0)$ featured below. Such is also the case for the commutativity associated with
the integral \cite{GT}

$$
\int_Cd^n\eta
d^n\eta'\sum_{N=0}^{N_0}G(N)Y_{\alpha}^N(\eta)Y_{\alpha}^N(\eta')$$
through the multiplicative factor $ {\mathcal L}^{(n)}(N_0)$
featured below in the result $ (1e)_5$. In what follows, for that
matter, it should be recalled that the factors $
\Lambda^{(n)}(N_0), L^{(n)}(N_0)$ and $ {\mathcal L}^{(n)}(N_0)$
result in constant, $ N'_0$-independent, expressions when
respectively multiplied with their associated finite sums over $
N'$ for each value of N. As a result, their exact form is
inconsequential to the renormalised effective action and to the
renormalisation of the theory itself.

The finite contribution to $ W_{I_r}$ at $ n \rightarrow 4$ stemming from the diagram in fig.1(c) is

$$
(1c):\hspace{0.5in} -\lambda^2ln(\mu a)\frac{(V_c+1)^2}{\pi^9}\frac{1}{2^{21}}RK^2\sum_{N=1}^{\infty}\frac{[C_N^{\frac{3}{2}}(cos\theta_4^0)]^2}{N^2(N+3)^2}\Lambda^{(4)}(N'_0)[\sum_{N'=0}^{N'_0}(2N'+3)]^2\int_Cd^4\eta
\hspace{0.7in}
$$
the finite contribution stemming from the diagram in fig.1(d) is

$$
(1d):\hspace{0.5in}
-\lambda^2\frac{1}{3}\frac{1}{2^{18}}\frac{V_c+1}{\pi}\gamma_ERK^2
\sum_{N=1}^{\infty}\frac{[C_N^{\frac{3}{2}}(cos{\theta_4^0})]^2}{N^2(N+3)^2}\Lambda^{(4)}(N'_0)\big{[}\sum_{N'=0}^{N'_0}(2N'+3)\big{]}^2\int_Cd^4\eta
\hspace{1.5in}
$$

$$
\hspace{0.3in}
+ \lambda^2ln(\mu a)\frac{1}{3}\frac{1}{2^{16}}\frac{V_c+1}{\pi}RK^2
\sum_{N=1}^{\infty}\frac{[C_N^{\frac{3}{2}}(cos{\theta_4^0})]^2}{N^2(N+3)^2}\Lambda^{(4)}(N'_0)\big{[}\sum_{N'=0}^{N'_0}(2N'+3)\big{]}^2\int_Cd^4\eta
$$
The contribution to $ W_{I_r}$ stemming from the diagram in fig.(1e) is particularly involved. As the derivation of that contribution is tedious, albeit straightforward, it will be given in some detail in the calculational context set in \cite{GT}. The symmetry factor of $ \frac{1}{48}$ is inherent in what follows.

The evaluation of the stated diagram is associated with five terms. The pole structure of the diagram in fig.1(e) relates, exclusively, to the first three terms and has been explicitly analyzed in \cite{GT}. The finite contribution stemming from the first term is
$$
(1e)_1:\hspace{0.3in}
-\lambda^2\frac{1}{\pi^{6}}\frac{1}{3^3}\frac{1}{2^{20}}\gamma_ERK^2\sum_{N=0}^{\infty}\big{[}
\frac{(N+2)(N+3)}{(N+1)(N+4)}
\big{(}C_{N+1}^{\frac{3}{2}}(cos{\theta_4^0})\big{)}^2\big{]}\int_Cd^4\eta
\hspace{0.7in}$$

$$
+
\lambda^2\frac{1}{\pi^{6}}\frac{1}{3^4}\frac{1}{2^{18}}ln(\mu a)RK^2\sum_{N=0}^{\infty}\big{[}
\frac{(N+2)(N+3)}{(N+1)(N+4)}
\big{(}C_{N+1}^{\frac{3}{2}}(cos{\theta_4^0})\big{)}^2\big{]}\int_Cd^4\eta $$
With $ \Omega_4$ associated with $ \partial C_4$ the finite contribution stemming from the second term is

$$
(1e)_2:\hspace{0.0in}+
\lambda^2\frac{1}{3^2}\frac{1}{2^{13}}\frac{1}{\pi^4}\gamma_ERK^2L^{(4)}(N'_0) \times $$

$$
\sum_{N=0}^{\infty}\sum_{N'=0}^{N'_0}\frac{(N+1)(N+2)}{[{N'}^2-N^2+3(N'-N)]^2}\frac{\Gamma(1+\frac{1}{N'_0})\Gamma(N'+1+\frac{1}{N'_0})}{\Gamma(N'+3+\frac{1}{N'_0})}
F^2N'(\Omega_5^{(C)})^{-1}\int_Cd^4{\eta} $$

$$
+
\lambda^2\frac{1}{3^2}\frac{1}{2^{12}}\frac{1}{\pi^4}\gamma_ERKL^{(4)}(N'_0) \times $$

$$
\sum_{N=0}^{\infty}\sum_{N'=0}^{N'_0}\frac{(N+1)(N+2)}{[{N'}^2-N^2+3(N'-N)]^2}\frac{\Gamma(1+\frac{1}{N'_0})\Gamma(N'+1+\frac{1}{N'_0})}{\Gamma(N'+3+\frac{1}{N'_0})}
[FBH]N'(\Omega_4)^{-1}(sin\theta_4^0)^{-3}\oint_{\partial C}d^3\eta $$

$$
- \lambda^2\frac{1}{3^3}\frac{1}{2^{15}}\frac{1}{\pi^4}\gamma_ER^2L^{(4)}(N'_0) \times $$

$$
\sum_{N=0}^{\infty}\sum_{N'=0}^{N'_0}\frac{(N+1)(N+2)}{[{N'}^2-N^2+3(N'-N)]^2}\frac{\Gamma(1+\frac{1}{N'_0})\Gamma(N'+1+\frac{1}{N'_0})}{\Gamma(N'+3+\frac{1}{N'_0})}
[BH]^2(\Omega_5^{(C)})^{-1}\int_Cd^4{\eta} $$

$$
- \lambda^2\frac{1}{3^2}\frac{1}{2^{12}}\frac{1}{\pi^4}ln(\mu a)RK^2L^{(4)}(N'_0)\times $$

$$
\sum_{N=0}^{\infty}\sum_{N'=0}^{N'_0}\frac{(N+1)(N+2)}{[{N'}^2-N^2+3(N'-N)]^2}\frac{\Gamma(1+\frac{1}{N'_0})\Gamma(N'+1+\frac{1}{N'_0})}{\Gamma(N'+3+\frac{1}{N'_0})}\times $$

$$
F^2N'(\Omega_5^{(C)})^{-1}\int_Cd^4{\eta} $$

$$
- \lambda^2\frac{1}{3^2}\frac{1}{2^{11}}\frac{1}{\pi^4}ln(\mu a)RKL^{(4)}(N'_0)\times $$

$$
\sum_{N=0}^{\infty}\sum_{N'=0}^{N'_0}\frac{(N+1)(N+2)}{[{N'}^2-N^2+3(N'-N)]^2}\frac{\Gamma(1+\frac{1}{N'_0})\Gamma(N'+1+\frac{1}{N'_0})}{\Gamma(N'+3+\frac{1}{N'_0})}\times $$

$$
[FBH]N'(\Omega_4)^{-1}(sin\theta_4^0)^{-3}\oint_{\partial C}d^3\eta $$

$$
+ \lambda^2\frac{1}{3^3}\frac{1}{2^{14}}\frac{1}{\pi^4}ln(\mu a)R^2L^{(4)}(N'_0)\times $$

$$
\sum_{N=0}^{\infty}\sum_{N'=0}^{N'_0}\frac{(N+1)(N+2)}{[{N'}^2-N^2+3(N'-N)]^2}\frac{\Gamma(1+\frac{1}{N'_0})\Gamma(N'+1+\frac{1}{N'_0})}{\Gamma(N'+3+\frac{1}{N'_0})}\times $$

$$
[BH]^2(\Omega_5^{(C)})^{-1}\int_Cd^4{\eta} $$
The finite contribution stemming from the third term is

$$
(1e)_3:\hspace{0.0in}
-\lambda^2\frac{1}{3}\frac{1}{2^{13}}\frac{1}{\pi^{4}}RK^2\gamma_EL^{(4)}(N'_0)\times $$

$$
\sum_{N=0}^{\infty}\sum_{N'=0}^{N'_0}
\frac{\Gamma\big{(}\frac{1}{N'_0}\big{)}}{\big{[}N'^2 - N^2 + 3(N' - N)\big{]}^2}
F^2N'\big{(}\Omega_5^{(C)}\big{)}^{-1}\int_Cd^4\eta
\hspace{0.5in}$$

$$
-\lambda^2\frac{1}{3}\frac{1}{2^{12}}\frac{1}{\pi^{4}}RK\gamma_EL^{(4)}(N'_0)\times $$

$$
\sum_{N=0}^{\infty}\sum_{N'=0}^{N'_0}
\frac{\Gamma\big{(}\frac{1}{N'_0}\big{)}}{\big{[}N'^2 - N^2 + 3(N' - N)\big{]}^2}
[FBH]N'\big{(}\Omega_4\big{)}^{-1}\big{(}sin\theta^0_4\big{)}^{-3}\oint_{\partial C}d^3\eta
\hspace{0.5in}$$

$$
-\lambda^2\frac{1}{3^2}\frac{1}{2^{15}}\frac{1}{\pi^{4}}R^2\gamma_EL^{(4)}(N'_0)\times $$

$$
\sum_{N=0}^{\infty}\sum_{N'=0}^{N'_0}
\frac{\Gamma\big{(}\frac{1}{N'_0}\big{)}}{\big{[}N'^2 - N^2 + 3(N' - N)\big{]}^2}
[BH]^2\big{(}\Omega_5^{(C)}\big{)}^{-1}\int_Cd^4\eta
\hspace{0.5in}$$

$$
+\lambda^2\frac{1}{3}\frac{1}{2^{11}}\frac{1}{\pi^{4}}ln(\mu a)RK^2L^{(4)}(N'_0)\times $$

$$
\sum_{N=0}^{\infty}\sum_{N'=0}^{N'_0}
\frac{\Gamma\big{(}\frac{1}{N'_0}\big{)}}{\big{[}N'^2 - N^2 + 3(N' - N)\big{]}^2}
F^2N'\big{(}\Omega_5^{(C)}\big{)}^{-1}\int_Cd^4\eta
\hspace{0.5in}$$

$$
+\lambda^2\frac{1}{3}\frac{1}{2^{10}}\frac{1}{\pi^{4}}ln(\mu a)RKL^{(4)}(N'_0)\times $$

$$
\sum_{N=0}^{\infty}\sum_{N'=0}^{N'_0}
\frac{\Gamma\big{(}\frac{1}{N'_0}\big{)}}{\big{[}N'^2 - N^2 + 3(N' - N)\big{]}^2}
[FBH]N'\big{(}\Omega_4\big{)}^{-1}\big{(}sin\theta^0_4\big{)}^{-3}\oint_{\partial C}d^3\eta
\hspace{0.5in}$$

$$
+\lambda^2\frac{1}{3^2}\frac{1}{2^{13}}\frac{1}{\pi^{4}}ln(\mu a)R^2L^{(4)}(N'_0)\times $$

$$
\sum_{N=0}^{\infty}\sum_{N'=0}^{N'_0}
\frac{\Gamma\big{(}\frac{1}{N'_0}\big{)}}{\big{[}N'^2 - N^2 + 3(N' - N)\big{]}^2}
[BH]^2\big{(}\Omega_5^{(C)}\big{)}^{-1}\int_Cd^4\eta
\hspace{0.5in}$$
The remaining two terms are finite at the dimensional limit $ \epsilon \rightarrow 0$. For that matter, they have not been pursued in \cite{GT} and \cite{George}. The present calculation, however, necessitates their explicit inclusion. The calculation of the fourth term advances on lines identical to those of the previous three terms. Its contribution is

$$
(1e)_4: +\lambda^2\frac{1}{\pi^4}\frac{1}{3^2}\frac{1}{2^{13}}RK^2L^{(4)}(N'_0)\times $$

$$
\sum_{N=0}^{\infty}\sum_{N'=0}^{N'_0}\frac{\Gamma(\frac{1}{N'_0}-1)\Gamma(N'+3+\frac{1}{N'_0})}{(N^2+3N+2)\big{[}N'^2 - N^2 + 3(N' - N)\big{]}^2\Gamma(N'+1+\frac{1}{N'_0})}\times
$$

$$
F^2N'\big{(}\Omega_5^{(C)}\big{)}^{-1}\int_Cd^4\eta
$$

$$
+\lambda^2\frac{1}{\pi^4}\frac{1}{3^2}\frac{1}{2^{12}}RKL^{(4)}(N'_0)\times $$

$$
\sum_{N=0}^{\infty}\sum_{N'=0}^{N'_0}\frac{\Gamma(\frac{1}{N'_0}-1)\Gamma(N'+3+\frac{1}{N'_0})}{(N^2+3N+2)\big{[}N'^2 - N^2 + 3(N' - N)\big{]}^2\Gamma(N'+1+\frac{1}{N'_0})}\times
$$

$$
[FBH]N'\big{(}\Omega_4\big{)}^{-1}\big{(}sin\theta_4^0\big{)}^{-3}\oint_{\partial C}d^3\eta
$$

$$
+\lambda^2\frac{1}{\pi^4}\frac{1}{3^3}\frac{1}{2^{15}}R^2L^{(4)}(N'_0)\times $$

$$
\sum_{N=0}^{\infty}\sum_{N'=0}^{N'_0}\frac{\Gamma(\frac{1}{N'_0}-1)\Gamma(N'+3+\frac{1}{N'_0})}{(N^2+3N+2)\big{[}N'^2 - N^2 + 3(N' - N)\big{]}^2\Gamma(N'+1+\frac{1}{N'_0})}\times
$$

$$
[BH]^2\big{(}\Omega_5^{(C)}\big{)}^{-1}\int_Cd^4\eta
$$
Finally, the fifth term amounts entirely to

$$
(1e)_5:
+ \lambda^2\frac{1}{\pi^6}\frac{1}{3^3}\frac{1}{2^{19}}RK^2{\mathcal L}^{(4)}(N_0)\times $$

$$
\sum_{N=0}^{N_0}\frac{\Gamma(-2 + \frac{1}{N_0})\Gamma(N + 4 + \frac{1}{N_0})}{N^2(N+3)^2\Gamma(N + \frac{1}{N_0})}\big{[}C_N^{\frac{3}{2}}(cos\theta_4^0)\big{]}^2(V_c + 1)\int_Cd^4\eta
$$

In these expressions use has been made of \cite{GT}, \cite{George}

$$
\int_Sd^4\eta = (V_c + 1)\int_Cd^4\eta ~~~; ~~~V_c > 0
$$
relating the volume of $ S_4$ to that of $ C_4$ as well as

$$
\oint_{\partial C} d^3\eta = a^3(sin\theta_4^0)^3\Omega_4
$$

The resulting finite contribution $ (1e)$ stemming from the diagram of fig.(1e) is, for that matter, the sum-total

$$
(1e) = (1e)_1 + (1e)_2 + (1e)_3 + (1e)_4 + (1e)_5
$$

\end{document}